\newcount\mgnf\newcount\tipi\newcount\tipoformule\newcount\greco 
\tipi=2          
\tipoformule=0   

\global\newcount\numsec\global\newcount\numfor
\global\newcount\numapp\global\newcount\numcap
\global\newcount\numfig\global\newcount\numpag
\global\newcount\numnf

\def\SIA #1,#2,#3 {\senondefinito{#1#2}%
\expandafter\xdef\csname #1#2\endcsname{#3}\else
\write16{???? ma #1,#2 e' gia' stato definito !!!!} \fi}

\def \FU(#1)#2{\SIA fu,#1,#2 }

\def\etichetta(#1){(\veroparagrafo.\veraformula)%
\SIA e,#1,(\veroparagrafo.\veraformula) %
\global\advance\numfor by 1%
\write15{\string\FU (#1){\equ(#1)}}%
\write16{ EQ #1 ==> \equ(#1)  }}
\def\etichettaa(#1){(A\veraappendice.\veraformula)
 \SIA e,#1,(A\veraappendice.\veraformula)
 \global\advance\numfor by 1
 \write15{\string\FU (#1){\equ(#1)}}
 \write16{ EQ #1 ==> \equ(#1) }}
\def\getichetta(#1){Fig. \verafigura
 \SIA g,#1,{\verafigura}
 \global\advance\numfig by 1
 \write15{\string\FU (#1){\graf(#1)}}
 \write16{ Fig. #1 ==> \graf(#1) }}
\def\retichetta(#1){\numpag=\pgn\SIA r,#1,{\verapagina}
 \write15{\string\FU (#1){\rif(#1)}}
 \write16{\rif(#1) ha simbolo  #1  }}
\def\etichettan(#1){(n\verocapitolo.\veranformula)
 \SIA e,#1,(n\verocapitolo.\veranformula)
 \global\advance\numnf by 1
\write16{\equ(#1) <= #1  }}

\newdimen\gwidth
\gdef\profonditastruttura{\dp\strutbox}
\def\senondefinito#1{\expandafter\ifx\csname#1\endcsname\relax}
\def\BOZZA{
\def\alato(##1){
 {\vtop to \profonditastruttura{\baselineskip
 \profonditastruttura\vss
 \rlap{\kern-\hsize\kern-1.2truecm{$\scriptstyle##1$}}}}}
\def\galato(##1){ \gwidth=\hsize \divide\gwidth by 2
 {\vtop to \profonditastruttura{\baselineskip
 \profonditastruttura\vss
 \rlap{\kern-\gwidth\kern-1.2truecm{$\scriptstyle##1$}}}}}
\def\verapagina{
{\romannumeral\number\numcap}.\number\numsec.\number\numpag}}

\def\alato(#1){}
\def\galato(#1){}
\def\veroparagrafo{\number\numsec}\def\veraformula{\number\numfor}
\def\veraappendice{\number\numapp}
\def\verapagina{\number\pageno}\def\veranformula{\number\numnf}
\def\verafigura{{\romannumeral\number\numcap}.\number\numfig}
\def\verocapitolo{\number\numcap}\def\veranformula{\number\numnf}
\def\Eqn(#1){\eqno{\etichettan(#1)\alato(#1)}}
\def\eqn(#1){\etichettan(#1)\alato(#1)}
\def\ver{\veroparagrafo}
\def\Eq(#1){\eqno{\etichetta(#1)\alato(#1)}}
\def\eq(#1){\etichetta(#1)\alato(#1)}
\def\Eqa(#1){\eqno{\etichettaa(#1)\alato(#1)}}
\def\eqa(#1){\etichettaa(#1)\alato(#1)}
\def\dgraf(#1){\getichetta(#1)\galato(#1)}
\def\drif(#1){\retichetta(#1)}

\def\eqv(#1){\senondefinito{fu#1}$\clubsuit$#1\else\csname fu#1\endcsname\fi}
\def\equ(#1){\senondefinito{e#1}\eqv(#1)\else\csname e#1\endcsname\fi}
\def\graf(#1){\senondefinito{g#1}\eqv(#1)\else\csname g#1\endcsname\fi}
\def\rif(#1){\senondefinito{r#1}\eqv(#1)\else\csname r#1\endcsname\fi}
\def\bib[#1]{[#1]\numpag=\pgn
\write13{\string[#1],\verapagina}}

\def\include#1{
\openin13=#1.aux \ifeof13 \relax \else
\input #1.aux \closein13 \fi}

\openin14=\jobname.aux \ifeof14 \relax \else
\input \jobname.aux \closein14 \fi
\openout15=\jobname.aux
\openout13=\jobname.bib

\let\EQ=\Eq

\ifnum\tipoformule=1\let\Eq=\eqno\def\eq{}\let\Eqa=\eqno\def\eqa{}
\def\equ{}\fi


{\count255=\time\divide\count255 by 60 \xdef\hourmin{\number\count255}
        \multiply\count255 by-60\advance\count255 by\time
   \xdef\hourmin{\hourmin:\ifnum\count255<10 0\fi\the\count255}}

\def\oramin{\hourmin }

\def\data{\number\day/\ifcase\month\or january \or february \or march \or
april \or may \or june \or july \or august \or september
\or october \or november \or december \fi/\number\year;\ \oramin}

\setbox200\hbox{$\scriptscriptstyle \data $}

\newcount\pgn \pgn=1
\def\foglio{\number\numsec:\number\pgn
\global\advance\pgn by 1}
\def\foglioa{A\number\numsec:\number\pgn
\global\advance\pgn by 1}

\footline={\rlap{\hbox{\copy200}}\hss\tenrm\folio\hss}

\def\TIPIO{
\font\setterm=amr7 
\def \settepunti{\def\rm{\fam0\setterm}
\textfont0=\setterm   
\normalbaselineskip=9pt\normalbaselines\rm
}\let\nota=\settepunti}

\def\TIPITOT{
\font\twelverm=cmr12
\font\twelvei=cmmi12
\font\twelvesy=cmsy10 scaled\magstep1
\font\twelveex=cmex10 scaled\magstep1
\font\twelveit=cmti12
\font\twelvett=cmtt12
\font\twelvebf=cmbx12
\font\twelvesl=cmsl12
\font\ninerm=cmr9
\font\ninesy=cmsy9
\font\eightrm=cmr8
\font\eighti=cmmi8
\font\eightsy=cmsy8
\font\eightbf=cmbx8
\font\eighttt=cmtt8
\font\eightsl=cmsl8
\font\eightit=cmti8
\font\sixrm=cmr6
\font\sixbf=cmbx6
\font\sixi=cmmi6
\font\sixsy=cmsy6
\font\twelvetruecmr=cmr10 scaled\magstep1
\font\twelvetruecmsy=cmsy10 scaled\magstep1
\font\tentruecmr=cmr10
\font\tentruecmsy=cmsy10
\font\eighttruecmr=cmr8
\font\eighttruecmsy=cmsy8
\font\seventruecmr=cmr7
\font\seventruecmsy=cmsy7
\font\sixtruecmr=cmr6
\font\sixtruecmsy=cmsy6
\font\fivetruecmr=cmr5
\font\fivetruecmsy=cmsy5
\textfont\truecmr=\tentruecmr
\scriptfont\truecmr=\seventruecmr
\scriptscriptfont\truecmr=\fivetruecmr
\textfont\truecmsy=\tentruecmsy
\scriptfont\truecmsy=\seventruecmsy
\scriptscriptfont\truecmr=\fivetruecmr
\scriptscriptfont\truecmsy=\fivetruecmsy
\def \eightpoint{\def\rm{\fam0\eightrm}
\textfont0=\eightrm \scriptfont0=\sixrm \scriptscriptfont0=\fiverm
\textfont1=\eighti \scriptfont1=\sixi   \scriptscriptfont1=\fivei
\textfont2=\eightsy \scriptfont2=\sixsy   \scriptscriptfont2=\fivesy
\textfont3=\tenex \scriptfont3=\tenex   \scriptscriptfont3=\tenex
\textfont\itfam=\eightit  \def\it{\fam\itfam\eightit}%
\textfont\slfam=\eightsl  \def\sl{\fam\slfam\eightsl}%
\textfont\ttfam=\eighttt  \def\tt{\fam\ttfam\eighttt}%
\textfont\bffam=\eightbf  \scriptfont\bffam=\sixbf
\scriptscriptfont\bffam=\fivebf  \def\bf{\fam\bffam\eightbf}%
\tt \ttglue=.5em plus.25em minus.15em
\setbox\strutbox=\hbox{\vrule height7pt depth2pt width0pt}%
\normalbaselineskip=9pt
\let\sc=\sixrm  \let\big=\eightbig  \normalbaselines\rm
\textfont\truecmr=\eighttruecmr
\scriptfont\truecmr=\sixtruecmr
\scriptscriptfont\truecmr=\fivetruecmr
\textfont\truecmsy=\eighttruecmsy
\scriptfont\truecmsy=\sixtruecmsy
}\let\nota=\eightpoint}

\newfam\msbfam   
\newfam\truecmr  
\newfam\truecmsy 
\newskip\ttglue
\ifnum\tipi=0\TIPIO \else\ifnum\tipi=1 \TIPI\else \TIPITOT\fi\fi

\global\newcount\numpunt

\magnification=\magstephalf
\baselineskip=16pt
\parskip=8pt

\def\a{\alpha}
\def\b{\beta}
\def\d{\delta}
\def\e{\epsilon}
\def\f{\phi}
\def\g{\gamma}
\def\k{\kappa}
\def\l{\lambda}
\def\r{\rho}
\def\s{\sigma}
\def\t{\tau}

\def\o{\omega}

\def\L{\Lambda}
\def\G{\Gamma}
\def\O{\Omega}

\def\del #1{\frac{\partial^{#1}}{\partial\l^{#1}}}

\def\E{{I\kern-.25em{E}}}
\def\N{{I\kern-.22em{N}}}
\def\M{{I\kern-.22em{M}}}
\def\R{{I\kern-.22em{R}}}
\def\Z{{Z\kern-.5em{Z}}}
\def\1{{1\kern-.25em\hbox{\rm I}}}
\def\eu{{1\kern-.25em\hbox{\sm I}}}
\def\f1{{1\kern-.25em\hbox{\vsm I}}}
\def\C{{C\kern-.75em{C}}}
\def\P{{I\kern-.25em{P}}}

\def\del{\partial}


\def\AA{{\cal A}}
\def\BB{{\cal B}}

\def\DD{{\cal D}}

\def\FF{{\cal F}}
\def\GG{{\cal G}}

\def\KK{{\cal K}}
\def\SS{{\cal S}}

\def\NN{{\cal N}}
\def\MM{{\cal M}}

\def\LL{{\cal L}}

\def\QQ{{\cal Q}}

\def\chap #1#2{\line{\ch #1\hfill}\numsec=#2\numfor=1}

\def\ba{{\backslash}}

\def\wt{\widetilde}

\def\limlaw{\buildrel \DD\over\rightarrow}

\newcount\foot
\foot=1
\def\note#1{\footnote{${}^{\number\foot}$}{\ftn #1}\advance\foot by 1}

\def\frac#1#2{{#1\over #2}}
\def\sfrac#1#2{{\textstyle{#1\over #2}}}
\def\text#1{\quad{\hbox{#1}}\quad}
\def\newpage{\vfill\eject}
\def\proposition #1{\noindent{\thbf Proposition #1:}}

\def\theo #1{\noindent{\thbf Theorem #1: }}
\def\lemma #1{\noindent{\thbf Lemma #1: }}

\def\corollary #1{\noindent{\thbf Corollary #1: }}
\def\proof{{\noindent\pr Proof: }}

\def\endproof{$\diamondsuit$}
\def\remark{\noindent{\bf Remark: }}
\def\thanks{\noindent{\bf Acknowledgements: }}
\font\pr=cmbxsl10
\font\thbf=cmbxsl10 scaled\magstephalf

\font\ch=cmbx12
\font\ftn=cmr8

\font\it=cmti10
\font\bf=cmbx10
\font\sm=cmr7
\font\vsm=cmr6


\overfullrule=0pt

\font\tit=cmbx12
\font\aut=cmbx12
\font\aff=cmsl12
\def\s{\char'31}
\nopagenumbers
{$  $}
\vskip1.5truecm
\centerline{\tit METASTATES IN THE HOPFIELD MODEL IN THE}
\vskip.2truecm
\centerline{\tit REPLICA SYMMETRIC REGIME\footnote{${}^\#$}{\ftn Work
partially supported by the Commission of the European Communities
under contract  CHRX-CT93-0411}}
\vskip1.5truecm
\centerline{\aut Anton Bovier 
\footnote{${}^1$}{\ftn e-mail:
bovier@wias-berlin.de}
}
\vskip.1truecm
\centerline{\aff Weierstra\s {}--Institut}
\centerline{\aff f\"ur Angewandte Analysis und Stochastik}
\centerline{\aff Mohrenstrasse 39, D-10117 Berlin, Germany}
\vskip.4truecm
\centerline{\aut  V\'eronique Gayrard\footnote{${}^2$}{\ftn
e-mail: gayrard@cpt.univ-mrs.fr}}
\centerline{\aff Centre de Physique Th\'eorique - CNRS}
\centerline{\aff Luminy, Case 907}
\centerline{\aff F-13288 Marseille Cedex 9, France}
\vskip1truecm\rm
\def\s{\sigma}
\noindent {\bf Abstract:}
We study the finite dimensional marginals of the Gibbs measure
in the Hopfield model at low temperature when the number of patterns, $M$,
is proportional to the volume with a sufficiently small proportionality 
constant $\a>0$. It is shown that even when a single pattern is selected 
(by a magnetic field or by conditioning), the marginals do not converge 
almost surely, but only in law. The corresponding limiting law is constructed 
explicitly. We fit our result in the recently proposed language of 
``metastates'' which we discuss in some length. As a byproduct,  in a certain 
regime of the parameters $\a$ and $\b$ (the inverse temperature),
 we also give a 
simple proof of Talagrand's [T1] recent result that the 
replica symmetric solution  found by Amit, Gutfreund, and Sompolinsky [AGS]
can be rigorously justified.


  \vfill
$ {} $

\newpage
\count0=1
\footline={\hss\tenrm\folio\hss}


\chap{1. Introduction}1

Strongly disordered systems such as spin glasses represent some of the most 
interesting and most difficult problems of statistical mechanics. Amongst the 
most remarkable achievements of  theoretical physics in this field is the 
exact solution of some models of mean field type via the replica trick and
Parisi's replica symmetry breaking scheme (For an exposition see [MPV];
the application to the Hopfield model [Ho] was carried out in [AGS]). 
The replica trick 
is a formal tool that allows to eliminate the difficulty of studying 
disordered systems by integrating out the randomness at the expense of
having to perform an analytic continuation of some function 
computable only on the positive integers to the value zero\note{As a 
matter of fact, such an analytic continuation is not performed. What is 
done is much more subtle: The function at integer values is represented as 
some integral suitable for evaluation by a saddle point method. Instead of 
doing this, apparently irrelevant critical points are selected judiciously 
and the ensuing wrong value of the function is then continued to the 
correct value at zero.}. Mathematically, this procedure is highly mysterious
and has so far resisted all attempts to be put on a solid basis. On the 
other hand, its apparent success is a clear sign that something ought to be 
understood better in this method. An apparently less mysterious approach that
yields the same answer is the cavity method [MPV].  However, 
here too, 
the derivation of the solutions involves a large number of intricate 
and unproven assumptions that seem hard or impossible to justify in general.

However, there has been some distinct progress in understanding the approach 
of the cavity method at least in simple cases where no breaking of
the replica symmetry occurs. The first attempts in this direction were made 
by   Pastur and Shcherbina [PS] in the Sherrington-Kirkpatrick model and 
Pastur, Shcherbina and Tirozzi [PST] in the Hopfield model. Their results
were conditional: They assert to show that the replica symmetric solution,
holds under certain unverified assumption, namely the vanishing of the 
so-called Edwards-Anderson parameter.  A breakthrough was achieved in a recent
paper by Talagrand [T1] where he proved the validity of the 
replica symmetric solution in an explicit domain of the model
parameters in the Hopfield model. His approach is purely by induction over
the volume  (i.e. the cavity method) and uses only some a priori estimates
on the support properties of the distribution of the so-called overlap 
parameters as first proven in [BGP1,BGP2] and in sharper form in [BG1].

Let us recall the definition of the Hopfield model
and some basic notations. 
Let $\SS_N\equiv \{-1,1\}^N$ denote the set  of functions
$\s:\{1,\dots,N\}\rightarrow \{-1,1\}$, and set $\SS\equiv \{-1,1\}^{\N}$.
We call $\s$ a spin configuration and denote by $\s_i$ the value of $\s$
at $i$. Let $(\O,\FF,\P)$ be an abstract probability space and let
$\xi^\mu_i$, $i,\mu\in \N$, denote a family of independent
identically distributed  random variables
on this space. For the purposes of this paper we will assume that
$\P[\xi_i^\mu=\pm 1]=\frac 12$. 
We will write $\xi^\mu[\o]$ for the $N$-dimensional random 
vector whose $i$-th
component is given by $\xi_i^\mu[\o]$
and call such a vector a `pattern'. On the other hand, we use the
notation
$\xi_i[\o]$ for the $M$-dimensional vector with the same components.
When we write $\xi[\o]$ without indices, we frequently will consider
it as an $M\times N$ matrix and we write $\xi^t[\o]$ for the transpose 
of this matrix. Thus, $\xi^t[\o]\xi[\o]$ is the $M\times M$ matrix
whose elements are $\sum_{i=1}^N\xi_i^\mu[\o]\xi_i^\nu[\o]$. With this
in mind we will use throughout the paper a vector notation with
$(\cdot,\cdot)$ standing for the scalar product in whatever space the
argument may lie. E.g. the expression $(y,\xi_i)$ stands for 
$\sum_{\mu=1}^M\xi_i^\mu y_\mu$, etc.

We define random maps $m_N^\mu[\o]:\SS_N\rightarrow [-1,1]$
through\note{We will make the dependence of random quantities on the
             random parameter $\o$ explicit by an added $[\o]$ whenever we want
             to stress it. Otherwise, we will frequently drop the
             reference to $\o$ to simplify the notation.}
$$
m_N^\mu[\o](\s)\equiv \frac 1N\sum_{i=1}^N \xi_i^\mu[\o]\s_i
\Eq(A.1)
$$
Naturally, these maps `compare' the configuration $\s$ globally to the
random configuration $\xi^\mu[\o]$. A Hamiltonian is now defined as the
simplest negative function of these variables, namely
$$
\eqalign{
H_N[\o](\s)&\equiv -\frac N2\sum_{\mu=1}^{M(N)} \left(m_N^\mu[\o](\s)\right)^2
\cr
& =-\frac N2 \left\|m_N[\o](\s)\right\|_2^2
}
\Eq(A.2)
$$
where $M(N)$ is some, generally increasing, function that crucially influences
the properties of the model. $\|\cdot\|_2$ denotes the
$\ell_2$-norm in $\R^M$, and the vector $m_N[\o](\s)$ is always understood
to be $M(N)$-dimensional.
 
Through this Hamiltonian we define in a natural way finite volume
Gibbs measures on $\SS_N$ via
$$
\mu_{N,\b}[\o](\s)\equiv \frac 1{Z_{N,\b}[\o]}e^{-\b H_N[\o](\s)}
\Eq(A.4)
$$
and the induced distribution of the overlap parameters
$$
\QQ_{N,\b}[\o]\equiv \mu_{N,\b}[\o]\circ m_N[\o]^{-1}
\Eq(A.5)
$$
The normalizing factor $Z_{N,\b}[\o]$, given by
$$
Z_{N,\b}[\o]\equiv 2^{-N}\sum_{\s\in \SS_N} e^{-\b H_N[\o](\s)}
\equiv \E_\s e^{-\b H_N[\o](\s)}
\Eq(A.6)
$$
is called the partition function. We are interested in the large $N$ behaviour
 of these measures. 
In our previous work we have been mostly concerned with the limiting 
induced measures. In this paper we return to the limiting behaviour
of the 
Gibbs measures themselves, making use, however, of the information 
obtained on the asymptotic properties of the  induced measures.

We pursue two objectives. Firstly, we give an alternative 
proof of Talagrand's result (with possibly a slightly different range of 
parameters) that, although equally based on the cavity method, makes 
more extensive use of the properties of the overlap-distribution that were 
proven   in [BG1]. This allows, in our opinion, some considerable 
simplifications. Secondly, we will elucidate some conceptual issues 
concerning the infinite volume Gibbs states in this model. Several delicacies 
in the question of convergence of finite volume Gibbs states 
(or local specifications) in highly disordered systems, and in particular
spin glasses, were pointed 
out repeatedly by Newman and Stein over the last years [NS1,NS2]. 
But only during the last year did they propose the formalism of so-called
``metastates'' [NS3,NS4,N] that seems to provide the appropriate 
framework to discuss these issues. In particular, we will show that in the
Hopfield model, this formalism seems unavoidable for spelling out 
convergence results.  

Let us formulate our main result in a slightly preliminary form (precise 
formulations require some more discussion and notation and will be given in 
Section 5). 

Denote by $m^*(\b)$ the largest solution of the mean field equation
$m=\tanh (\b m)$ and  by $e^\mu$ the $\mu$-th unit vector of 
the canonical basis 
of $\R^M$.  For all $(\mu,s)\in\{-1,1\}\times\{1,\dots,M\}$
let $B_{\rho}^{(\mu,s)}\subset\R^M$ denote the ball of radius $\rho$
centered at $ s m^* e^\mu$.
For any pair of indices $(\mu, s)$ and any $\rho>0$ we define the  
conditional measures
$$
\mu_{N,\b,\rho}^{(\mu,s)}[\o](\AA)\equiv 
\mu_{N,\b}[\o]( \AA\mid B_{\rho}^{(\mu,s)}),\,\,\,\,\,\,\AA\in\BB(\{-1,1\}^{N})
\Eq(A.7)
$$
\def\mm{m_1}
\def\rr{r}
\def\qq{q}
The so called ``replica symmetric equations''\note{We cite these equations, 
(3.3-5) in [AGS] only for the case $k=1$, where $k$ is the number
of the so-called ``condensed patterns''. One could generalize our
results presumably measures conditioned on balls around ``mixed
states'', i.e. the metastable states with more than one ``condensed pattern'',
but we have not worked out the details.}
 of [AGS] is the following system of equations in three unknowns 
$\mm,\rr$, and $\qq$, given by
$$
\eqalign{
\mm&=\int d\NN(g) \tanh(\b(\mm+\sqrt{\a\rr} g))\cr
\qq&=\int d\NN(g) \tanh^2(\b(\mm+\sqrt{\a\rr} g))\cr
\rr&=\frac{\qq}{(1-\b+\b\qq)^2}
}
\Eq(A.7bis)
$$
With this notation we can state

\theo {\ver.1} {\it There exists a nonempty connected set of parameters 
$\b,\a$ bounded by the curves $\a=0$, $\a= c (m^*(\b))^4$ and
$\b= c' \a$, such that if $\lim_{N\uparrow \infty}M(N)/N=\a$
the following holds: 
For any finite $I\subset\N$, and for any $s_{I}\subset \{-1,1\}^I$, 
$$
\mu_{N,\b,\rho}^{(\mu,s)}(\{\s_I=s_I\})\rightarrow
\prod_{i\in I}\frac {e^{\b s_i\left[\mm\xi_i^1+g_i\sqrt {\a\rr}\right]}}
{2\cos\left(\b\left[\mm\xi_i^1
+g_i\sqrt {\a\rr}\right]\right)}
\Eq(A.8)
$$
as $N\uparrow\infty$, where the $g_i$, $i\in I$ are independent gaussian 
random variables with mean zero and variance one that are independent of
the random variables $\xi_i^1$, $i\in I$. The convergence is understood in law
with respect to the distribution of the gaussian variables $g_i$.
}

This theorem should be juxtaposed to our second result:

\theo{\ver.2} {\it On the same set of parameter as in Theorem \ver.1, the 
following is true with probability one:
For any finite $I\subset \N$ and for any $x\in \R^I$, there exist 
subsequences $N_k[\o]\uparrow\infty$ such that  
for any $s_{I}\subset \{-1,1\}^I$, if $\a >0$,
$$
\lim_{k\uparrow\infty}\mu_{N_k[\o],\b,\rho}^{(\mu,s)}[\o](\{\s_I=s_I\})=
\prod_{i\in I}\frac {e^{s_ix_i}}
{2\cosh( x_i)}
\Eq(A.9)
$$
}

The above statements may look a little bit surprising and need clarification.
This will be the main purpose of Section 2, where  we 
give a rather detailed discussion of the problem of 
convergence and the notion of metastates with the particular issues in 
disordered mean field models in view. We will also propose yet a  different 
notion of a state (let us call it ``superstate''), that tries to capture the 
asymptotic volume dependence of Gibbs states in the form of a continuous time
measure valued stochastic process. We also discuss the issue of 
the ``boundary conditions'' or rather ``external fields'', and the 
construction of conditional Gibbs measures in this context. This will 
hopefully prepare the ground for the understanding of our results in the 
Hopfield case. 

The following two section collect technical preliminaries. Section 3
recalls some results on the overlap distribution from [BG1-3] that will
be crucially needed later. Section 4 states and proves a version of the 
Brascamp-Lieb
inequalities [BL] that is suitable for our situation. 

Section 5 contains our central results. Here we construct explicitly the 
finite dimensional marginals of the Gibbs measures in finite volume and
study their behaviour in the infinite volume limit. The results will
be stated in the language of metastates. In this section we assume the 
convergence of certain thermodynamic functions which will be proven in
Section 6. Modulo this, this section contains the precise statements
and proofs of Theorems \ver.1 and \ver.2.

In Section 6 we give a proof of the convergence of these quantities and we 
relate them to the replica symmetric solution. This sections is largely 
based on the ideas of [PST]  and [T1] and is mainly added for the 
convenience of the reader.

\thanks We gratefully acknowledge helpful discussions on metastates
with Ch. Newman and Ch. K\"ulske.

\newpage

\chap{2. Notions of convergence of random Gibbs measures.}2

In this section we make some remarks on the appropriate picture 
for the study of limiting Gibbs measures for disordered systems,
with particular regard to the situation in mean-field like systems.
Although some of the observations we will make here arose naturally 
from the properties we discovered in the Hopfield model, our understanding 
has been greatly enhanced by the recent work of Newman and Stein [NS3,NS4,N]
and their introduction of the concept of ``metastates''. We refer the reader 
to their papers for more detail and further applications. Some nice examples 
can also be found in [K,BGK]. Otherwise, we keep this section self-contained
and geared for the situation we will describe in the Hopfield model, 
although part of the discussion is very general and not restricted to mean
field situations. For this reason we talk about finite volume measures
indexed by finite sets $\L$ rather then by the integer $N$.

\noindent {\bf Metastates.} 
The basic objects of study are {\it finite volume Gibbs measures},
$\mu_{\L,\b}$ (which for convenience we will always consider as 
measures on the infinite product space $\SS_\infty$).
We denote by 
$\left(\MM_1(\SS_\infty),\GG\right)$ the measurable space 
of probability measures on $\SS_\infty$ equipped with the sigma-algebra
$\GG$ generated by the open sets with respect to the weak topology on 
$\MM_1(\SS_\infty)$\note{Note that a basis of  open sets is given
by  sets of the 
forms
$\scriptstyle
\NN_{f_1,\dots,f_k,\e}(\mu)\equiv \{\mu'|\forall_{1\leq i\leq k} |\mu(f_i)
-\mu'(f_i)|<\e\}$, where $\scriptstyle f_i$ are continuous functions on 
$\scriptstyle\SS^\infty$;
indeed, it is enough to consider cylinder functions.}.
We will always regard Gibbs measures as random variables on the underlying 
probability space $(\O,\FF,\P)$ with values in the space $\MM_1(\SS_\infty)$,
 i.e. as measurable maps
$\O\rightarrow \MM_1(\SS_\infty)$. 

We are in principle interested in considering weak limits of these measures
as $\L\uparrow\infty$. There are essentially three things that may happen:

\item{(1)} Almost sure convergence:  For $\P$-almost all $\o$,
$$
\mu_{\L}[\o]\rightarrow \mu_\infty[\o]
\Eq(C.1)
$$
where $\mu_\infty[\o]$ may or may not depend on $\o$ (in general it will).
\item{(2)} Convergence in law: 
$$
\mu_{\L}\limlaw \mu_\infty
\Eq(C.2)
$$
\item{(3)} Almost sure convergence along random subsequences: 
There exist (at least for almost all $\o$) subsequences $\L_i[\o]\uparrow\infty$ such that 
$$
\mu_{\L_i[\o]}[\o]\rightarrow \mu_{\infty,\{\L_i[\o]\}}[\o]
\Eq(C.3)
$$

In systems with compact single site state space, (3) holds always, 
and there are models with non-compact state space where it holds with the 
``almost sure'' provision (see e.g. [BK]). 
 However, this contains little
information, if the subsequences along which convergence holds are only known 
implicitly. In particular, it gives no information on how, for any given large 
$\L$ the measure $\mu_\L$ ``looks like approximately''. In contrast, if 
(i) holds, we are in a very nice situation, as for any large enough $\L$
and for (almost) any realization of the disorder, the 
measure $\mu_{\L}[\o]$ is well approximated by $ \mu_\infty[\o]$. Thus, the 
situation would be essentially like in an ordered system (the ``almost sure''
excepted). It seems to us that the common feeling of most people 
working in the field of disordered systems was that this could be arranged
by putting suitable boundary conditions or external fields, to 
``extract pure states''. Newman and Stein [NS1] were, to our knowledge,
 the first to 
point to difficulties with this point of view. 
{\it In fact, there is no reason why
we should ever be, or be able to put us,  in a situation where (1) holds}, 
and this possibility should be considered as perfectly exceptional.
With (3) uninteresting and (1) unlikely, we are left with (2). By
 compactness, (2) holds always at least for (non-random!) subsequences $\L_n$,
 and even convergence without 
subsequences can be expected rather commonly. On the other hand, 
(2) gives us very reasonable information on our system, telling us what is the 
chance that our measure $\mu_\L$ for large $\L$ will look like some measure
$\mu_\infty$. This is much more than what (3) tells us, and baring the 
case where (1) holds, all we may reasonably expect to know. 

We should thus investigate the case (2) more closely. As proposed actually 
first by Aizenman and Wehr [AW], it is most natural to consider an object 
$K_\L$ defined as a measure on the product space $\O\otimes \MM_1(\SS_\infty)$
(equipped with the product topology and the weak topology, respectively), such
  that its marginal distribution on $\O$ is $\P$ while 
the conditional measure, $\k_\L(\cdot)[\o]$,
on $\MM_1(\SS_\infty)$ given $\FF$\note{We write shorthand $\scriptstyle\FF$
for $\scriptstyle\MM_1(\SS^\infty)\otimes\FF$ whenever appropriate.}  
is the Dirac measure on 
$\mu_\L[\o]$; the marginal on $\MM_1(\SS_\infty)$ is then of course the 
law of $\mu_\L$. The advantage of this construction over simply regarding 
the law of $\mu_\L$ lies in the fact that we can in this way extract more 
information by conditioning, as we shall explain. 
Note that by compactness $K_\L$ converges at least along (non-random!)
subsequences, and we may assume that it actually 
converges to some measure $K$. Now the case (1) above corresponds to the 
situation where the conditional probability on $\GG$ given $\FF$
is degenerate, i.e.   
$$
\k(\cdot)[\o]=\d_{\mu_\infty[\o]}(\cdot), \text{ a.s.}
\Eq(C.4)
$$
Thus we see that in general even the conditional distribution
$\k(\cdot)[\o]$ of 
$K$ is a nontrivial measure on the space of infinite volume Gibbs measures,
this latter object being called the (Aizenman-Wehr) metastate\note{It may be 
interesting to recall the reasons that led Aizenman and Wehr to this 
construction. In their analysis of the effect of quenched diorder on phase 
transition they required the existence of ``translation-covariant'' states.
Such object could be constructed as weak limits of finite volume
states with e.g. periodic or translation invariant boundary conditions,
provided the corresponding sequences converge almost surely (and not via
subsequences with possibly different limits). They noted that in a general 
disordered system this may not be true. The metastate provided a way out of 
this difficulty.}. 
What happens is
that the asymptotic properties of the Gibbs measures as the volume tends to 
infinity depend in a intrinsic way on the tail sigma field of the 
disorder variables, and even after all random variables are fixed, some 
``new'' randomness appears that allows only probabilistic statements on the 
asymptotic Gibbs state. 

\noindent{\it A toy example:} {It may be useful to illustrate the passage from
convergence in law to the Aizenman-Wehr metastate in a more familiar context,
namely the ordinary central limit theorem.
Let $(\O,\FF, \P)$ be a probability space, and let 
$\{X_i\}_{i\in \N}$ be a family of i.i.d. centered 
random variables with variance one; let $\FF_n$ be the
sigma algebra generated by $X_1,\dots, X_n$ and let 
$\FF\equiv \lim_{n\uparrow \infty}\FF_n$. Define the real valued random 
variable
$G_n\equiv \frac 1{\sqrt n}\sum_{i=1}^n X_i$. We may define the joint law
$K_n$ of $G_n$ and the $X_i$ as a probability measure on 
$\R\otimes \O$. Clearly, this measure converges to some measure
$K$ whose marginal on $\R$ will be the standard normal distribution. However,
we can say more, namely

\noindent{\bf Toy-Lemma {\ver.1}} {\it In the example described above, the conditional measure 
$\k(\cdot)[\o]\equiv K(\cdot|\FF)$ satisfies
$$
\k(\cdot)[\o]=\NN(0,1),\text{$\P$-a.s.}
\Eq(C.4bis)
$$
}

\proof We need to understand what \eqv(C.4bis) means. Let $f$ be a continuous 
function on $\R$. We claim that for almost all $\o$, 
$$
\int f(x)\k(dx)[\o] =\int \frac {e^{-x^2/2}}{\sqrt{2\pi}}f(x)dx
\Eq(C.4ter)
$$
Define the martingale $h_n\equiv\int f(x) K(dx,d\o|\FF_n)$. We may write
$$
\eqalign{
h_n&=\lim_{N\uparrow\infty}\E_{X_{n+1}}\dots\E_{X_N}f\left(\sfrac 1{\sqrt N}
\sum_{i=1}^N X_i\right) \cr
&=\lim_{N\uparrow\infty}\E_{X_{n+1}}\dots\E_{X_N}f\left(\sfrac 1{\sqrt {N-n}}
\sum_{i=n+1}^N X_i\right),\text{a.s.} \cr
&=\int \frac {e^{-x^2/2}}{\sqrt{2\pi}}f(x)dx, 
}
\Eq(C.4quater)
$$
where we used that for fixed $N$, $\frac 1{\sqrt N}\sum_{i=1}^nX_i$ 
converges to 
zero as $N\uparrow \infty$ almost surely.
Thus, for any continuous $f$, $h_n$ is almost surely  constant, while
$\lim_{n\uparrow\infty} h_n =\int f(x) K(dx,d\o|\FF)$, by the martingale 
convergence theorem. This proves the lemma.
\endproof
}

The CLT example may inspire the question whether one might not be able to 
retain more information on the convergence of the random Gibbs state
than is kept in the Aizenman-Wehr metastate. The metastate tells us about the
probability distribution of the limiting measure, but we have thrown out
all information on how for a given $\o$, the finite volume measures behave as
the volume increases. 

 Newman and Stein [NS3,NS4] have introduced a possibly more profound 
concept of the 
{\it empirical metastate} which captures more precisely the 
asymptotic volume dependence of the Gibbs states in the infinite volume 
limit. We will briefly discuss this object and elucidate its meaning in the 
above CLT context. Let $\L_n$ be an increasing and absorbing sequence of 
finite volumes. Define the random empirical measures $\k_N^{em}(\cdot)[\o]$
on $\left(\MM_1(\SS^\infty)\right)$ by
$$
\k_N^{em}(\cdot)[\o]\equiv \frac 1N\sum_{n=1}^N \d_{\mu_{\L_n}[\o]}
\Eq(C.10)
$$
In [NS4] it was proven that for sufficiently sparse sequences
$\L_n$ and subsequences $N_i$, it is true that
almost surely 
$$
\lim_{i\uparrow\infty} \k_{N_i}^{em}(\cdot)[\o]
=\k(\cdot)[\o]
\Eq(C.11)
$$
Newman and Stein conjectured that in many 
situations, the use of sparse subsequences 
would not be necessary to achieve the above convergence. However,
K\"ulske [K] has exhibited some simple mean field examples 
where almost sure convergence only holds for very sparse (exponentially
spaced) subsequences). He also showed that for more slowly growing 
sequences convergence in law can be proven in these cases.

\noindent {\it Toy example revisited:} {All this is easily understood in our example. We set
$G_n \equiv \frac 1{\sqrt n}\sum_{i=1}^n X_i$. Then the empirical 
metastate corresponds to 
$$
\k_N^{em}(\cdot)[\o]\equiv \frac 1N\sum_{n=1}^N \d_{G_n[\o]}
\Eq(C.12)
$$

We will prove that the following Lemma holds:

\noindent{\bf Toy-Lemma} {\ver.2} {\it Let $G_n$ and 
$\k_N^{em}(\cdot)[\o]$ be defined above.
Let $B_t$, $t\in [0,1]$ denote a standard Brownian motion. Then

\item{(i)} The random measures $\k_N^{em}$ converge in law to the 
measure $\k^{em}=\int_0^1 dt \d_{t^{-1/2}  B_t}$
\item{(ii)} 
$$
\E\left[ \k^{em}(\cdot)|\FF\right] 
= \NN(0,1)
\Eq(C.13)
$$
}

\proof Our main objective is to prove (i). We will see that quite clearly,
this result relates to Lemma \ver.1 as the CLT to the Invariance Principle,
and indeed, its proof is essentially an immediate consequence of 
Donsker's Theorem.   Donsker's theorem (see [HH] for a formulation
in more generality than needed in this chapter) asserts the following:
Let 
$\eta_n(t)$ denote the continuous function on $[0,1]$ that 
for $t=k/n$ is given by
$$
\eta_n(k/n)\equiv \frac 1{\sqrt n}\sum_{i=1}^k X_i
\Eq(C.14)
$$
and that interpolates linearly between these values for all other points $t$.
Then, $\eta_n(t)$ converges in distribution to standard Brownian
motion in the sense that for any continuous functional 
$F:C([0,1])\rightarrow \R$ it is true that $F(\eta_n)$ converges in law
to 
$F(B)$. 
From here the proof of (i) is obvious. We have to proof that for any bounded
continuous function $f$,
 $$
\eqalign{
& \frac 1N\sum_{n=1}^N \d_{G_n[\o]}(f)
\equiv \frac 1N\sum_{n=1}^N  f\left(\eta_n(n/N)/\sqrt{n/N}\right )
\rightarrow \cr
&\int_0^1 dt f(B_t/\sqrt t)\equiv  
\int_0^1 dt \d_{B_t/\sqrt t}(f)
}
\Eq(C.15)
$$
To see this, simply define the continuous functionals $F$ and $F_N$ by
$$
F(\eta)\equiv \int_0^1 dt f(\eta(t)/\sqrt t)
\Eq(C.16)
$$
and 
$$
F_N(\eta)\equiv \frac 1N\sum_{n=1}^N f(\eta(n/N)/\sqrt{n/N})
\Eq(C.17)
$$
We have to show that in distribution $F(B)-F_N(\eta_N)$ converges to zero.
But 
$$
F(B)-F_N(\eta_N)=F(B)-F(\eta_N)+F(\eta_N)-F_N(\eta_N)
\Eq(C.18)
$$
By the invariance principle, $F(B)-F(\eta_N)$ converges to zero in 
distribution while $F(\eta_N )-F_N(\eta_N)$ converges to zero 
since $F_N$ is the Riemann sum approximation to $F$. 

To see that (ii) holds, note first that as in the CLT, the brownian motion 
$B_t$ is measurable with respect to the tail sigma-algebra of the $X_i$. Thus
$$
\E\left[\k^{em}|\FF\right] =\NN(0,1)
\Eq(C.19)
$$
\endproof

\remark It is easily seen that 
for sufficiently sparse subsequences $n_i$ (e.g. $n_i=i!$),
$$
\frac 1N\sum_{i=1}^N \d_{G_{n_i}}\rightarrow \NN(0,1), \text {a.s}
\Eq(C.191)
$$
but the weak convergence result contains 
in a way more information. 
}

\noindent{\bf Superstates:} In our example we have seen that 
the empirical metastate converges in distribution to the empirical measure
of the stochastic process $B_t/\sqrt t$. It appears natural   
to think that the construction of the corresponding continuous time 
stochastic process itself is actually the right way to look at the problem 
also in the context of random Gibbs measures, and that the 
 the empirical metastate could 
converge (in law) to the empirical measure of this 
process. To do this we propose the following, yet somewhat tentative 
construction.
  
We fix again  a sequence of finite volumes $\L_n$\note{The outcome of our 
construction will depend on the choice of this sequence. Our philosophy here
would be to choose a natural sequence of volumes for the problem at hand.
In mean field examples this would be 
$\scriptstyle\L_n=\{1,\dots,n\}$, on a lattice one might choose cubes of 
sidelength $n$.}. We define
for $t\in [0,1]$
$$
\mu^t_{\L_n}[\o]\equiv (t-[tn]/n)\mu_{\L_{[tn]+1}}[\o] +
(1-t+[tn]/n)\mu_{\L_{[tn]}}[\o] 
\Eq(C.20)
$$ 
(where as usual $[x]$ denote the smallest integer less than or equal to $x$).
Clearly this object is a continuous time 
stochastic process whose state space is $\MM_1(\SS)$. We may try to construct 
the limiting process
$$
\mu_t[\o]\equiv \lim_{n\uparrow\infty} \mu^t_{\L_n}[\o]
\Eq(C.21)
$$
where the limit again can in general be expected only 
in distribution.  Obviously, in our CLT example, 
this is precisely how we construct the Brownian motion in the invariance 
principle.   
We can now of course repeat the construction of the Aizenman-Wehr metastate
on the level of processes. 
To do this, one must make some choices for the topological 
space one wants to work in. A natural possibility is to consider
the space
$
C\left([0,1],\MM_1(\SS^\infty)\right)
$ of continuous measure valued function equipped with the 
uniform weak topology\note{Another possibility would be a measure valued 
version of the space $\scriptstyle D([0,1],\MM_1(\SS))$ of
measure valued  C\`adl\`ag functions. 
The choice depends essentially on the properties we expect from
the limiting process (i.e. continuous sample paths or not).}, 
i.e. we say that a sequence of its 
elements $\l_i$ converges to $\l$, if and only if, 
for all continuous functions 
$f:\SS^\infty\rightarrow\R$, 
$$
\lim_{i\rightarrow\infty} \sup_{t\in [0,1]} \left|\l_{i,t}(f)-\l_t(f)\right|
=0
\Eq(C.22)
$$
Since the weak topology is metrizable, so is the uniform weak topology 
and $C\left([0,1],\MM_1(\SS^\infty)\right)$ becomes a metric space so we may
 define
the corresponding sigma-algebra generated by the open sets. Taking the 
tensor product with our old $\O$,  we can 
thus introduce the set    
$\MM_1\left(C\left([0,1],\MM_1(\SS^\infty)\right)\otimes
\O\right)$
of probability measures on this space tensored with $\O$. 
Then we define the elements 
$$
\KK_n\in \MM_1\left(C\left([0,1],\MM_1(\SS^\infty)\right)\otimes\O\right)
$$
whose marginals on $\O$ are $\P$ and whose conditional measure on 
$C\left([0,1],\MM_1(\SS^\infty)\right)$, given $\FF$ are the Dirac measure on 
the measure valued function 
$\mu_{\L_{[tn]}}[\o]$, $t\in [0,1]$. 
Convergence, and even the existence of limit points for this sequence of
measures is now no longer a trivial matter. The problem of the existence 
of limit points can be circumvented by using a weaker notion of convergence,
e.g. that of the convergence of any finite dimensional marginal. Otherwise,
some tightness condition  is needed [HH], e.g.
we must check that for any continuous function $f$,
$\sup_{|s-t|\leq \d} |\mu^t_{\L_n}(f)-\mu^s_{\L_n}(f)|$ converges to zero in 
probability, uniformly in $N$, as $\d\downarrow 0$.\note {There are
pathological examples in which we would not expect such a result to be true.
An example is the ``highly disordered spin glass model'' of Newman and Stein
[NS5]. Of course, tightness may also be destroyed by choosing very rapidly 
growing sequences of volumes $\scriptstyle\L_n$.}

We can always hope that the limit 
as $n$ goes to infinity of $\KK_n$ exists. {\it If} the limit, $\KK$ exists, 
we can again consider its
conditional distribution given $\FF$, and the resulting object is the 
functional analog of the Aizenman-Wehr metastate. (We feel tempted to call 
this object the ``superstate''. Note that the marginal distribution of the 
superstate ``at time $t=1$'' is the Aizenman-Wehr metastate, and the 
law of the empirical distribution of the underlying process is the
empirical metastate). The ``superstate''  contains an enormous 
amount of information 
on the asymptotic volume dependence of the random Gibbs measures; on the 
other hand, its construction in any explicit form is generally hardly feasible.
  
Finally, we want to stress that the superstate will normally depend on
the choice of the basic sequences $\L_n$ used in its
construction. This feature
is already present in the empirical metastate. In particular, 
sequences growing extremely fast will give different results than
slowly increasing sequences. On the other hand, the very precise
choice
of the sequences should not be important. A natural choice would
appear to us sequences of cubes of sidelength $n$, or, in mean field
models,
simply the sequence of volumes of size $n$.

\noindent{\bf Boundary conditions, external fields, conditioning.}
In the discussion of Newman and Stein, metastates are usually constructed 
with simple boundary conditions such as periodic or ``free'' ones. They 
emphasize the feature of the ``selection of the states'' by the disorder
in a given volume without any bias through boundary conditions or symmetry 
breaking  fields. Our point of view is somewhat different in this respect
in that we think that the idea to apply special boundary conditions 
or, in mean field models, symmetry breaking terms, to 
improve convergence properties,  is still to some extend useful, the aim
ideally being to achieve the situation (1). Our only restriction in this is
really that our procedure shall have some {\it predictive power},
that is, it should give information of the approximate form of 
a finite volume Gibbs state. This excludes any construction
involving subsequences via compactness arguments. We thus are interested to 
know to what extend it is possible to reduce the ``choice'' of available 
states for the randomness to select from, to  smaller subsets and to classify
the minimal possible subsets (which then somehow 
play the r\^ole of {\it extremal 
states}). 
In fact, in the examples considered in [K,BGK] it would be possible
to reduce the size of such subsets to one, 
while in the example of the present paper,
we shall see that this is {\it impossible}. We have to discuss this point 
carefully.

While in short range  lattice models the DLR construction gives a clear 
framework  how the class of infinite volume Gibbs measures  is to be defined,
in mean field models this situation is somewhat ambiguous and needs discussion.

If the infinite volume 
Gibbs measure is unique (for given $\o$), quasi by definition, (1) 
must hold. So our problems arise from non-uniqueness. Hence the 
following recipe: modify $\mu_\L$ in such a way that uniqueness holds, 
while otherwise perturbing it in a minimal way. Two procedures suggest themselves:
\item{(i)} Tilting, and
\item{(ii)} Conditioning

Tilting consists in the addition of a {\it symmetry breaking } term to the 
Hamiltonian whose strength is taken to zero. Mostly, this term is taken 
{\it linear} so that it has the natural interpretation of a {\it magnetic
field}. More precisely,
define
$$
\mu_{\L,\e}^{\{h\}}[\o](\cdot)\equiv \frac {\mu_\L[\o]\left(\cdot \,
e^{-\b\e \sum_{i\in\L}h_i\s_i} \right)}
{\mu_\L[\o]\left(
e^{-\b\e \sum_{i\in\L}h_i\s_i} \right)}
\Eq(C.5)
$$
Here $h_i$ is some sequence of numbers that in general will have to be 
allowed to depend on $\o$ if anything is to be gained. 
One may also allow them to 
depend on $\L$ explicitly, if so desired. From a physical point of view 
we might wish to add further conditions, like some locality of the 
$\o$-dependence; in principle there should be a way of writing them down in 
some explicit way. 
We should stress that tilting by linear functions is not always satisfactory, 
as some states that one might wish to obtain are lost; an example is the
generalized Curie-Weiss model with Hamiltonian $H_N(\s) = -\frac N4
[m_N(\s)]^4$ at the critical point. There, the free energy has 
three degenerate absolute minima at $-m^*,0$, and $+m^*$, and while 
we might want to think of tree coexisting phases, only the measures centered
at $\pm m^*$ can be extracted by the above method. Of course this can be 
remedied by allowing arbitrary perturbation $h(m)$  with the only
condition that $\|h\|_\infty$ tends to zero at the end.

By conditioning we mean always conditioning  the macroscopic variables 
to be in some set $\AA$. This appears natural
since, in lattice models, extremal measures can always be extracted from 
arbitrary DLR measures by conditioning on events in the tail sigma fields;
the macroscopic variables are measurable with respect to the tail sigma 
fields.
Of course only conditioning on events that do not have too small
 probability will be reasonable. Without going into too much of a motivating 
discussion, we will adopt the following conventions. Let $\AA$ be an event 
in the sigma algebra generated by the macroscopic function. Put
$$
f_{\L,\b}(\AA)=-\frac 1{\b|\L|}\ln \mu_{\L,\b}[\o](\AA)
\Eq(C.6)
$$
We call $\AA$ admissible for conditioning if and only if 
$$
\lim_{|\L|\uparrow\infty} f_{\L,\b}[\o](\AA)=0
\Eq(C.7)
$$
We call $\AA$ minimal if it cannot be decomposed into 
two admissible subsets. In analogy with \eqv(C.5) we then define
$$
\mu^{\AA}_{\L,\b}[\o](\cdot)\equiv \mu_{\L,\b}[\o]\left(\cdot|\AA\right)
\Eq(C.8)
$$
We define the set of all limiting  Gibbs measures to be the set of limit points
of measures $\mu^{\AA}_{\L,\b}$ with admissible sets $\AA$. Choosing $\AA$ 
minimal, we improve our chances of obtaining convergent sequences
and the resulting limits are  serious candidates 
for {\it extremal} limiting Gibbs measures, but 
we stress that this is not guaranteed to succeed, 
as will become manifest in our
examples.
 This will not
mean that adding such conditioning is not going to be useful. 
It is in fact, as it 
will reduce the disorder in the metastate and may in general allow to 
construct various {\it different} metastates in the case of phase transitions.
The point to be understood here is that within the general framework 
outlined above, we should consider two different notions of {\it uniqueness}:

\item{(a)} {\it Strong uniqueness} meaning that for almost all $\o$ 
there is only one limit point $\mu_\infty[\o]$, and
\item{(b)} {\it Weak uniqueness\note{Maybe the notion of meta-uniqueness 
would be more appropriate}} meaning that there is a unique 
metastate, in the sense that for any choice of $\AA$, the 
metastate constructed taking the infinite volume limit with the measures
 $\mu_{\L,\e}^{\AA}$ is the same.

In fact, it may happen that the addition of a symmetry breaking term or 
conditioning  
does not lead to strong uniqueness. Rather, what may be true is that such 
a field selects a subset of the states, but to which of them
the state at given volume resembles can depend on the volume in a 
complicated way.

If weak uniqueness does not hold, one has a non-trivial set of metastates. 

It is quite clear that a sufficiently general tilting approach is equivalent 
to the conditioning approach; we prefer for technical reasons to use the
conditioning in the present paper. We also note that by dropping condition
\eqv(C.7) one can enlarge the class of limiting measures obtainable to 
include {\it metastable states}, which in many applications, in particular 
in the context of dynamics, are also relevant.

\newpage

\chap{3. Properties of the induced measures.}3

In this section we collect a number of results on the 
distribution of the overlap parameters in the Hopfield model
that were obtained in some of our previous papers [BG1,BG2,BG3].
We cite these results mostly from [BG3] where they were stated in the most
suitable form for our present purposes and we refer  the reader to that 
paper for the proofs.

We recall some notation. 
Let  $m^*(\b)$ be the largest solution of the mean field equation
$m=\tanh (\b m)$. Note that $m^*(\b)$ is strictly positive for all 
$\b>1$, $\lim_{\b\uparrow\infty}m^*(\b)=1$, $\lim_{\b\downarrow 1}
\frac{(m^*(\b))^2}{3(\b-1)}=1$ and $m^*(\b)=0$ if $\b\leq 1$. 
Denoting by $e^\mu$ the $\mu$-th unit vector of the canonical basis 
of $\R^M$ we set, for all $(\mu,s)\in\{-1,1\}\times\{1,\dots,M(N)\}$,
$$
m^{(\mu,s)}\equiv s m^*(\b)e^\mu, 
\Eq(1.7)
$$
and for any $\rho>0$ we define the balls 
$$
B_{\rho}^{(\mu,s)}\equiv
\left\{x\in\R^M \big| \|x-m^{(\mu,s)}\|_2\leq\rho\right\}
\Eq(1.8)
$$
For any pair of indices $(\mu, s)$ and any $\rho>0$ we define the  
conditional measures
$$
\mu_{N,\b,\rho}^{(\mu,s)}[\o](\AA)\equiv 
\mu_{N,\b}[\o]( \AA\mid B_{\rho}^{(\mu,s)}),\,\,\,\,\,\,\AA\in\BB(\{-1,1\}^{N})
\Eq(1.10)
$$
and the corresponding induced measures
$$
\QQ_{N,\b,\rho}^{(\mu,s)}[\o](\AA)\equiv 
\QQ_{N,\b}[\o]( \AA\mid B_{\rho}^{(\mu,s)}),\,\,\,\,\,\,\AA\in\BB(\R^{M(N)})
\Eq(1.11)
$$
The point here is that for $\rho\geq c\frac {\sqrt{\a}}{m^*(\b)}$, the 
sets $B_\rho^{(\mu,s)}$ are admissible in the sense of the last section. 

It will be extremely useful to introduce the Hubbard-Stratonovich 
transformed measures $\wt\QQ_{N,\b}[\o]$ which are nothing but 
the convolutions of the induced measures with a gaussian measure of
mean zero and 
variance $1/\b N$, i.e.
$$
\wt \QQ_{N,\b}[\o]\equiv \QQ_{N,\b}[\o]\star \NN(0,\frac {\1}{\b N})
\Eq(1.11bis)
$$
Similarly we define the conditional Hubbard-Stratonovich transformed measures
$$
\wt \QQ_{N,\b,\rho}^{(\mu,s)}[\o](\AA)\equiv 
\wt\QQ_{N,\b}[\o]( \AA\mid B_{\rho}^{(\mu,s)}),\,\,\,\,\,\,\AA\in\BB(\R^{M(N)})
\Eq(1.12)
$$
We will need to consider the Laplace transforms of these measures which we will
denote by\note{This notation is  slightly different from the one used 
in [BG3].}
$$
\LL_{N,\b,\rho}^{(\mu,s)}[\o](t)\equiv
\int e^{(t,x)}
d\QQ_{N,\b,\rho}^{(\mu,s)}[\o](x)
\,,\,\,\,\,\,\,\,
t\in\R^{M(N)} 
\Eq(1.13)
$$
and
$$
\wt\LL_{N,\b,\rho}^{(\mu,s)}[\o](t)\equiv
\int e^{(t,x)}
d\wt\QQ_{N,\b,\rho}^{(\mu,s)}[\o](x)
\,,\,\,\,\,\,\,\,
t\in\R^{M(N)} 
\Eq(1.14)
$$
The following is a simple adaptation of Proposition 2.1 of [BG3] to these
notations. 

\proposition{\ver.1} {\it Assume that $\b>1$. There exist finite 
positive  constants 
$c\equiv c(\b), \tilde c\equiv\tilde c(\b), \bar c\equiv \bar c(\b)$ 
such that, with  probability one, for all but a finite number of 
indices 
$N$, if $\rho$ satisfies
$$
\frac{1}{2}m^*
>\rho>
c(\b)\left\{\sfrac{1}{N^{1/4}}\wedge\sqrt{\a}\right\}
\Eq(1.15)
$$
then,
for all $t$ with $\frac{\|t\|_2}{\sqrt N}<\infty$,
\item{i)}  
$$
\LL_{\b,N,\rho}^{(\mu,s)}[\o](t)
\left(1-e^{-\tilde c M}\right)
\leq
e^{-\frac{1}{2N\b}\|t\|_2^2}\wt\LL_{\b,N,\rho}^{(\mu,s)}[\o](t)
\leq
e^{- \tilde c M}+\LL_{\b,N,\rho}^{(\mu,s)}(t)
\left(1+e^{- \tilde c M}\right)
\Eq(1.16)
$$
\item{ii)} for any  $\rho,\bar\rho$ satisfying \eqv(1.15) 
$$
\wt\LL_{\b,N,\bar\rho}^{(\mu,s)}[\o](t)
\left(1-e^{-\bar c M}\right)
\leq
\wt\LL_{\b,N,\rho}^{(\mu,s)}[\o](t)
\leq
e^{- \bar c M}+\wt\LL_{\b,N,\bar\rho}^{(\mu,s)}[\o](t)
\left(1+e^{- \bar c M}\right)
\Eq(2.17)
$$
\item{iii)}   for any  $\rho,\bar\rho$ satisfying \eqv(1.15)  
$$
\left|\left(\int d\QQ_{N,\b,\rho}^{(\mu,s)}[\o](m) m -
\int d\wt\QQ^{(\mu,s)}_{N,\b,\bar\rho}[\o](z)z\,\,,t\right)
\right|\leq   \|t\|_2 e^{-\bar c M}
\Eq(2.031)
$$
}

A closely related result that we will need is also an adaptation 
of estimates from [BG3], i.e. it is obtained combining 
Lemmata 3.2 and 3.4 of that paper.

\lemma{\ver.2}{\it There exists $\g_a>0$, such that 
for all $\b>1$ and $\sqrt{\a}<\g_a(m^*)^2$, if 
$c_0\frac{\sqrt{\a}}{m^*}<\rho< m^*/\sqrt 2$ then, with  probability 
one, for all but a finite number of indices $N$, 
for all $\mu\in\{1,\dots,M(N)\}$, $s\in\{-1,1\}$,
 for all 
$b>0$ such that $\rho+b<\sqrt 2 m^*$,
$$
1\leq\frac
{\QQ_{\b, N}\left(B_{\rho+b}^{(\mu,s)}\right)}
{\QQ_{\b, N}\left(B_{\rho}^{(\mu,s)}\right)}
\leq 1+e^{-c_2\b M}
\Eq(2.3bis)
$$
where $0<c_2<\infty$ is a numerical constant.
}

We finally recall our result on local convexity of the 
function $\Phi$.

\theo{\ver.3} {\it Assume that $1<\b<\infty$. 
If the parameters $\a,\b,\rho$ are such that for $\e>0$, 
$$
\eqalign{
\inf_{\t}&\Bigl(\b(1-\tanh^2(\b m^*(1-\t)))(1+3\sqrt\a)\cr
&+
2\b \tanh^2(\b m^*(1-\t))\G(\a,\t m^*/\rho)\Bigr)\leq 1-\e
}\Eq(LL.19zero)
$$
Then with probability one for all 
but a finite number of indices $N$, $\Phi_{N,\b}[\o](m^* e^1+v)$ is a twice  
differentiable and strictly convex function of $v$ on the set $\{v:\|v\|_2\leq
 \rho\}$, and  
$$
\l_{min}\left(\nabla^2 \Phi_{N,\b}[\o](m^* e^1+v)\right)>\e
\Eq(LL.191)
$$
on this set.}

\remark
This theorem was first obtained in [BG1], the above form is cited and proven 
in [BG2]. With $\rho$ chosen as  $\rho=c \frac {\sqrt\a} {m^*}$, the condition 
\eqv(LL.19zero) means 
(i) For $\b$ close to $1$: $\frac {\sqrt{\a}}{(m^*)^2}$ small and,
(ii) For $\b$ large:  $\a\leq c\b^{-1}$.
The condition on $\a$ for large $\b$ seems unsatisfactory, but one may 
easily convince oneself  that it cannot be substantially improved. 

\newpage

\def\div{\,\hbox{div}}
\chap{4. Brascamp-Lieb inequalities.}4

A basic tool of our analysis are the so-called Brascamp-Lieb inequalities [BL].
In fact, we need such inequalities in  a slightly different setting than 
they are presented in the literature, namely for measures with bounded support
on some domain $D\subset\R^M$. 
Our derivation follows the one given  in [H]
(see also [HS]), and is in this context almost obvious.

 Let $D\subset \R^M$ be a bounded connected domain. 
Let $V\in C^2(D)$ be a twice 
continuously differentiable function on $D$, 
let $\nabla^2 V$ denote its Hessian 
matrix and
assume that, for all $x\in D$,
$\nabla^2 V(x)\geq c>0$ (where we say that a matrix $A>c$, if and only if
for all $v\in R^M$, $(v,Av)\geq c (v,v)$). We define the probability measure
$\nu$ on $(D,\BB(D))$ by
$$
\nu(dx)\equiv \frac {e^{-NV(x)}d^Mx}{\int_De^{-NV(x)}d^Mx}
\Eq(X.1)
$$
Our central result is

\theo{\ver.1}{\it Let $\nu$ the probability measure defined above.
Assume that $f,g\in C^{1}(D)$, and assume that (w.r.g.) $\int_D d\nu(x)g(x)=
\int_D d\nu(x) f(x)=0$. Then 
$$
\eqalign{
\left|\int_D d\nu(x) f(x)g(x)\right|&\leq \frac 1{cN} \int_D d\nu(x)
\left\|\nabla f(x)\|_2\|\nabla g(x)\right\|_2\cr
&+\frac 1{cN}\frac{\int_{\del D}|g(x)|\left\|\nabla f(x)\right\|_2 
e^{-NV(x)}d^{M-1}x}
{\int_D e^{-NV(x)}d^Mx}
}
\Eq(X.2)
$$
where $d^{M-1}x$ is the Lebesgue measure on $\del D$. }

\proof We consider the Hilbert space $L^2(D,\R^M,\nu)$
of $R^M$ valued functions on $D$ with scalar product $\langle F,G\rangle
\equiv
\int_Dd\nu(x)(F(x),G(x))$. Let $\nabla$ be the 
gradient operator on $D$ defined with a domain of all 
bounded $C^1$-function that vanish on $\del D$. Let $\nabla^*$ denote its 
adjoint. Note that 
$\nabla^*=-e^{NV(x)}\nabla e^{-NV(x)}=-\nabla +N(\nabla V(x))$. One easily 
verifies by partial integration that on this domain the operator
$\nabla\nabla^*\equiv \nabla e^{NV(x)}\nabla e^{-NV(x)}=\nabla^*\nabla
+N\nabla^2V(x)$ is symmetric
and $\nabla^*\nabla\geq 0$, so that by our hypothesis,
$\nabla\nabla^*\geq cN>0$. As a consequence,
$\nabla\nabla^* $ has a self-adjoint extension whose inverse 
$(\nabla\nabla^*)^{-1}$ exists on all  $L^2(D,\R^M,\nu)$                       and is bounded in norm by $(cN)^{-1}$. 

As a consequence of the above, for any $f\in C^1(D)$, we can uniquely 
solve  the 
differential equation
$$
\nabla\nabla^*\,\nabla u =\nabla f
\Eq(X.3)
$$
for $\nabla u$. Now note that \eqv(X.3) implies that 
$\nabla^*\nabla u = f+ k$, where $k$ is a constant\note{Observe that this 
is only true because $\scriptstyle D$ is connected. For $\scriptstyle D$ 
consisting of several 
connected components the theorem is obviously false.}. Hence
for real valued  $f$ and $g$ as in the statement of the theorem,
$$
\eqalign{
\int_Dd\nu(x)\left(\nabla g(x),\nabla u(x)\right)     
&=\int_D d\nu(x) e^{NV(x)}\,\div\left(e^{-NV(x)}g\nabla u(x)\right)
+ \int_Dd\nu(x) g(x)\nabla^*\nabla u(x)\cr
&=\frac 1Z\int_{D} d^Mx\,\div\left(e^{-NV(x)}g\nabla u(x)\right)
+\int_Dd\nu(x) g(x)f(x)
}
\Eq(X.4)
$$
where $Z\equiv \int_D d^Mx\,e^{-NV(x)}$.
Therefore, taking into account that $\nabla u =(\nabla\nabla^*)^{-1}\nabla f$,
$$
\eqalign{
\left|\int_Dd\nu(x) g(x)f(x) \right|
&\leq \left|\int_Dd\nu(x)\left(\nabla g(x),
(\nabla\nabla^*)^{-1}\nabla f(x)\right) \right|\cr
&+\frac 1Z\left|
\int_{D} d^Mx\,\div\left(e^{-NV(x)}g\nabla u(x)\right)\right|\cr
&\leq  \frac 1{cN} \int_Dd\nu(x)\|\nabla g(x)\|_2
\|\nabla f(x)\|_2 \cr
&+
\frac 1{cNZ}\int_{\del D}|g(x)|\left\|\nabla f(x)\right\|_2 
e^{-NV(x)}d^{M-1}x
}
\Eq(X.4bis)
$$
Note that in  second term we used the Gauss-Green formula to convert the 
integral over a divergence into a surface integral.
This concludes the proof.\endproof

\remark As is obvious from the proof above and as was pointed out 
in [H], one can replace the bound on the lowest eigenvalue of the Hessian
of $V$
by a bound on the lowest eigenvalue of the operator $\nabla \nabla^*$.
So far we have not seen how to get a better bound on this eigenvalue in 
our situation, but it may well be that this observation can be a clue to an
improvement of our results. 

The typical situation where we want to use Theorem \ver.1 is the following:
Suppose we are given a measure like \eqv(X.1) but not on $D$, but on some
bigger domain. We may be able to establish the lower bound on $\nabla^2 V$
not everywhere, but only on the smaller domain $D$, but such that 
the measure is essentially concentrated on $D$ anyhow. It is then likely that
we can also estimate away the boundary term in \eqv(X.2), either because 
$V(x)$ will be large on $\del D$, or because $\del D$ will be very small
(or both). We then have essentially the Brascamp-Lieb inequalities at
our disposal.

We mention the following corollary which shows that the Brascamp-Lieb 
inequalities  give rise to concentration inequalities under certain conditions.

\corollary {\ver.2} {\it Let $\nu$ be as in Lemma \ver.3. Assume that 
$f\in C^1(D)$ and that moreover
$V_t(x)\equiv
 V(x)-tf(x)/N$ for $t\in [0,1]$ is still 
strictly convex and $\l_{min}(\nabla^2 V_t)\geq c'>0$. Then
$$
\eqalign{
0\leq&\ln \int_Dd\nu(x) e^{f(x)} -\int_D d\nu(x) f(x)\leq \frac 1{2c' N}
\sup_{t\in [0,1]} \int_D d\nu_t(x)\|\nabla f\|_2^2 \cr
&+\sup_{t\in [0,1]} \frac 1{c'N}\frac{\int_{\del D}|g(x)|\left\|\nabla f(x)\right\|_2 
e^{-NV_t(x)}d^{M-1}x}
{\int_D e^{-NV_t(x)}d^Mx}
}
\Eq(X.5)
$$
where $\nu_t$ is the corresponding measure with $V$ replaced by $V_t$.
}

\proof Note that
$$
\eqalign{
\ln \E_V e^{f}&=\E_V f+\int_{0}^1 ds\int_0^s d s'\frac{
\E_V\left[e^{s'f}\left(f-
\frac{\E_Ve^{s'f}f}
{\E_Ve^{s'f}}\right)^2\right]}
{\E_Ve^{s'f}}\cr
&
=\E_V f+\int_{0}^1 ds\int_0^s d s' \E_{V_{s'}} 
\left(f-\E_{V_{s'}}f\right)^2
}
\Eq(X.6)
$$
where by assumption
$V_s(x)$ has the same properties as $V$ itself.
Thus using \eqv(X.2) gives \eqv(X.6).\endproof

\remark We would like to note that a concentration estimate like 
Corollary \ver.2
can also be derived under slightly different hypothesis on $f$ using 
logarithmic Sobolev inequalities (see [Le]) 
which hold under the same hypothesis as Theorem
\ver.1, and which in fact can be derived as a special case using 
$f=h^2$ and $g=\ln h^2$ in Theorem \ver.1. 

In the situations where we will apply the Brascamp-Lieb inequalities,
the correction terms due to the finite domain $D$ will be totally irrelevant. 
This follows from the following simple observation.

\lemma {\ver.3} {\it Let $B_\rho$ denote the ball of radius $\rho$ 
centered at the origin. 
Assume that for all $x\in D$,
$d\geq \nabla^2 V(x)\geq c>0$. 
If $x^*$ denotes the unique minimum of $V$, assume 
that $\|x^*\|_2\leq \rho/2$. Then there exists a constant $K<\infty$ 
(depending only on $c$ and $d$) such that if $\rho\geq K\sqrt{M/N}$, then
for $N$ large enough
$$
\frac{\int_{\del D} 
e^{-NV(x)}d^{M-1}x}
{\int_D e^{-NV(x)}d^Mx}\leq e^{- \rho^2 N/K}
\Eq(X.7)
$$
}

The proof of this lemma is elementary and will be left to the reader.

\newpage

\def\EQ{\E_{\Phi_{N}}}
\def\EQX{\E_{\Phi_N,x}}
\chap{5. The convergence of the Gibbs measures.}5

After these preliminaries we can now come to the central part of the 
paper, namely the study of the marginal distributions of the Gibbs measures
$\mu_{N,\b,\rho}^{(\mu,s)}$. Without loss of generality it suffices to 
consider the case $(\mu,s)=(1,1)$, of course.
 Let us fix $I\subset \N$ arbitrary but 
finite. We assume that $\L\supset I$, and for notational simplicity 
we put $|\L|=N+|I|$. We are interested in the probabilities
$$
\mu^{(1,1)}_{\L,\b,\r}[\o]\left(\{\s_I=s_I\}\right)
\equiv \frac{\E_{\s_{\L\ba I}}e^{\frac 12\b |\L| 
\left\|m_\L(s_I,\s_{\L\ba I})\right\|_2^2}\1_{\{m_\L(s_I,\s_{\L\ba I})
\in B_\rho^{(1,1)}\}}}
{\E_{\s_I}\E_{\s_{\L\ba I}}e^{\frac 12\b |\L| 
\left\|m_\L(\s_I,\s_{\L\ba I})\right\|_2^2}\1_{\{m_\L(s_I,\s_{\L\ba I})
\in B_\rho^{(1,1)}\}} }
\Eq(G.1)
$$
Note that $\|m_I(\s)\|_2\leq \sqrt M$. 
Now we can write
$$
m_\L(\s) =\frac{N}{|\L|} m_{\L\ba I}(\s)+\frac {|I|}{|\L|}m_I(\s)
\Eq(G.2)
$$ 
Then 
$$
\eqalign{
&\1_{\{m_\L(s_I,\s_{\L\ba I})
\in B_\rho^{(1,1)}\}}\leq \1_{\{  m_{\L\ba I}(\s)\in 
 B_{\rho_+}^{(1,1)}\}}\cr
&\1_{\{m_\L(s_I,\s_{\L\ba I})
\in B_\rho^{(1,1)}\}}\geq \1_{\{  m_{\L\ba I}(\s)\in 
 B_{\rho_- }^{(1,1)}\}}\cr
}
\Eq(G.2bis)
$$
where $\rho_\pm\equiv \rho\pm\frac{\sqrt M |I|}N$.
Setting $\b'\equiv \frac {N}{|\L|}\b$, this allows us to write
$$
\eqalign{
\mu^{(1,1)}_{\L,\b,\r}[\o]&\left(\{\s_I=s_I\}\right)\leq
\frac {\int_{B_{\rho_+}^{(1,1)}}d\QQ_{\L\ba I,\b'}(m)
e^{\b'|I|(m_I(s_I),m)} \, e^{\b \frac {|I|^2}{2|\L|}\|m_I(s_I)\|_2^2}}
{2^{|I|}\E_{\s_I}\int_{B_{\rho_-}^{(1,1)}}d\QQ_{\L\ba I,\b'}(m)
e^{\b'|I|(m_I(\s_I),m)} \, e^{\b \frac {|I|^2}{2|\L|}\|m_I(\s_I)\|_2^2}}
\cr
&\times \frac {\int_{B_{\rho_-}^{(1,1)}}d\QQ_{\L\ba I,\b'}(m)}
{\int_{B_{\rho_+}^{(1,1)}}d\QQ_{\L\ba I,\b'}(m)}\cr
&\leq 
\frac {\LL_{\L/I,\b,\rho_+}[\o](\b'|I| m_I(s_I))\,  
e^{\b \frac {|I|^2}{2|\L|}\|m_I(s_I)\|_2^2}}
{2^{|I|}\E_{\s_I} \LL_{\L/I,\b,\rho_-}[\o](\b'|I| m_I(\s_I))\,  
e^{\b \frac {|I|^2}{2|\L|}\|m_I(\s_I)\|_2^2}}
 \frac {\QQ_{\L\ba I,\b'}\left(B_{\rho_+}^{(1,1)}      \right)}
{\QQ_{\L\ba I,\b'}\left(B_{\rho_-}^{(1,1)} \right)}
}
\Eq(G.2ter)
$$
and 
$$
\eqalign{
\mu^{(1,1)}_{\L,\b,\r}[\o]&\left(\{\s_I=s_I\}\right)\geq
\frac {\int_{B_{\rho_-}^{(1,1)}}d\QQ_{\L\ba I,\b'}(m)
e^{\b'|I|(m_I(s_I),m)} \, e^{\b \frac {|I|^2}{2|\L|}\|m_I(s_I)\|_2^2}}
{2^{|I|}\E_{\s_I}\int_{B_{\rho_+}^{(1,1)}}d\QQ_{\L\ba I,\b'}(m)
e^{\b'|I|(m_I(\s_I),m)} \, e^{\b \frac {|I|^2}{2|\L|}\|m_I(\s_I)\|_2^2}}
\cr
&\times \frac {\QQ_{\L\ba I,\b'}\left(B_{\rho_-}^{(1,1)}      \right)}
{\QQ_{\L\ba I,\b'}\left(B_{\rho_+}^{(1,1)} \right)}\cr
&=
\frac {\LL_{\L/I,\b,\rho_-}[\o](\b'|I| m_I(s_I))\,  
e^{\b \frac {|I|^2}{2|\L|}\|m_I(s_I)\|_2^2}}
{2^{|I|}\E_{\s_I} \LL_{\L/I,\b,\rho_+}[\o](\b'|I| m_I(\s_I))\,   
e^{\b \frac {|I|^2}{2  |\L|}\|m_I(\s_I)\|_2^2}}
\frac {\QQ_{\L\ba I,\b'}\left(B_{\rho_-}^{(1,1)}      \right)}
{\QQ_{\L\ba I,\b'}\left(B_{\rho_+}^{(1,1)} \right)}
}
\Eq(G.2quater)
$$

Now  the term $\frac {|I|^2}{ N}\|m_I(s)\|_2^2$ is, up to a constant
that is independent of the $s_i$, irrelevantly small. More precisely, we 
have that

\lemma{\ver.1}{\it There exist $\infty>C,c>0$ such that for 
all $I$, $M$, and for
all $x>0$,
$$
\eqalign{
&\P\left[\sup_{\s_I\in\{-1,1\}^I}\sfrac {|I|^2}N \left|\|m_I(s)\|_2^2
-\sfrac {M|I|}N\right|\geq \sfrac {|I|M}N\left(\sqrt{\sfrac {|I|}N}+x\right)
\right]\cr
&\leq C\exp\left(- cM \left(\sqrt{1+x}-1\right)^2\right)
}
\Eq(G.7)
$$
}

\proof This Lemma is a direct consequence of estimates on the norm of the 
random matrices obtained, e.g. in Theorem 4.1 of [BG6].\endproof

Together with Proposition 3.1 and Lemma 3.2, we can now extract the 
desired representation for our probabilities.

\lemma {\ver.2} {\it For all $\b>1$ and $\sqrt{\a}<\g_a(m^*)^2$, if 
$c_0\frac{\sqrt{\a}}{m^*}<\rho< m^*/\sqrt 2$ then, with  probability 
one, for all but a finite number of indices $N$, 
for all $\mu\in\{1,\dots,M(N)\}$, $s\in\{-1,1\}$,
\item{(i)}
$$
\eqalign{
\mu^{(1,1)}_{\L,\b,\r}[\o]\left(\{\s_I=s_I\}\right)&=
\frac {\LL^{(1,1)}_{\L/I,\b,\rho}[\o](\b'|I| m_I(s_I))}
{2^{|I|}\E_{\s_I} \LL^{(1,1)}_{\L/I,\b,\rho}[\o](\b'|I| m_I(\s_I))}  
\cr
&+O(N^{-1/4})
}       
\Eq(G.8)
$$
and alternatively
\item{(ii)}
$$
\eqalign{
\mu^{(1,1)}_{\L,\b,\r}[\o]\left(\{\s_I=s_I\}\right)&=
\frac {\wt\LL^{(1,1)}_{\L/I,\b,\rho}[\o](\b'|I| m_I(s_I))}
{2^{|I|}\E_{\s_I} \wt\LL^{(1,1)}_{\L/I,\b,\rho}[\o](\b'|I| m_I(\s_I))}  
\cr
&+ O\left(e^{-O\left(M\right)}\right)
}
\Eq(G.9)
$$
}

We leave the details of the proof to the reader. We see that the 
computation of the marginal distribution of the Gibbs measures requires 
nothing but the computation of the Laplace transforms of the induced measures
 or
its Hubbard-Stratonovich transform at the random points $t=\sum_{i\in I}
s_i\xi_i$. Alternatively, these can be seen as the Laplace transforms of the 
distribution of the random variables $(\xi_i,m)$.

Now it is physically very natural that the law of the random variables
$(\xi_i,m)$ should determine the Gibbs measures completely. 
The point is that in a mean field model, the distribution 
of the spins in a finite set $I$ is determined
entirely in terms of the effective mean fields produced by the rest of the 
system that act on the spins $\s_i$. These fields are precisely 
the $(\xi_i,m)$. In a ``normal'' mean field situation, 
the mean fields are constant almost surely with respect to the
Gibbs measure. In the Hopfield model with subextensively many patterns, this 
will also be true, as $m$ will be concentrated near one of the values 
$m^* e^\mu$ (see [BGP1]). In that case $(\xi_i,m)$ will 
depend only in a local and very explicit form on the disorder, and the Gibbs 
measures will inherit this property. In a more general situation, 
the local mean fields may have a more complicated distribution, in particular
they may not be constant under the Gibbs measure, and 
the question is how to determine this. The approach of the {\it cavity method}
(see e.g. [MPV]) as carried out by Talagrand [T1] consists in deriving this 
distribution by induction over the volume. [PST] also followed this 
approach, using however the assumption of ``self-averaging'' of the 
order parameter to control errors. Our approach consists in 
using the detailed knowledge obtained on the measures $\wt\QQ$, and in 
particular the local convexity to determine a priori the form of the 
distribution; induction will then only be used to determine the remaining
few parameters.   
 
Let us begin with some general preparatory steps which will not yet 
require special properties of our measures. To simplify the notation, we 
we introduce the following abbreviations:

We write $\E_{\Phi_{N}}$ for the 
expectation with respect to the measures $ \wt\QQ_{\L\ba I,\b,h}[\o]$
conditioned on $B_\rho$  
and we set
$\bar Z\equiv Z-\E_{\Phi_{N}} Z$. 
We will write $\E_{\xi_I}$ for the expectation with respect to the 
family of random variables $\xi_i^\mu$, $i\in I$, $\mu=1,\dots,M$.

The first step in the computation of our Laplace transform consists in 
centering, i.e. we write
$$
\EQ e^{\sum_{i\in I}\b s_i (\xi_i,Z)}=
e^{\sum_{i\in I}\b s_i (\xi_i,\EQ Z)} 
\EQ e^{\sum_{i\in I}\b s_i (\xi_i,\bar Z)}
\Eq(G.10)
$$
While the  first factor will be entirely responsible 
for the for the distribution of the spins, our main efforts have to go into
controlling the second. 
To do this we will use heavily the fact, established first in [BG1],
that on $B_\rho^{(1,1)}$ the function $\Phi$ is convex with probability 
close to one. This allows us to exploit the Brascamp-Lieb inequalities in 
the form given in Section 3. The advantage of this procedure is that
it allows us to identify immediately the leading terms and to get a priori
estimates on the errors. This is to be contrasted to the much more involved
procedure of Talagrand [T1] who controls the errors by induction.

\noindent{\bf General Assumption:} For the remainder of this paper we will
always assume that the parameters $\a$ and $\b$ of our model are such that the
hypotheses of Proposition 3.1 and Theorem 3.3 are satisfied. All lemmata, 
propositions and theorem are valid under this provision only. 

\lemma {\ver.3} { \it Under our general assumption,
\item{(i)}
$$
\E_{\xi_I}\EQ e^{\sum_{i\in I}\b s_i (\xi_i,\bar Z)}
=e^{\frac {\b^2}2\sum_{i\in I} s_i^2 \EQ \|\bar Z\|_2^2}
\times e^{O(1/(\e N))}
\Eq(G.11)
$$
\item{(ii)} There is a finite constant $C$ such that 
$$
\E_{\xi_I}\left[\ln\left(\frac {\EQ e^{\sum_{i\in I}\b s_i (\xi_i,\bar Z)}}
{\E_{\xi_I} \EQ e^{\sum_{i\in I}\b s_i (\xi_i,\bar Z)}}\right)
\right]^2\leq \frac CN
\Eq(G.12)
$$
}

\remark The immediate consequence of this lemma is the observation that
the family of random variables $\left\{(\xi_i,\bar Z)\right\}_{i\in I}$ 
is asymptotically 
close to a family of i.i.d. centered gaussian 
random variables with variance $U_N\equiv\EQ \|\bar Z\|_2^2$. $U_N$ will be
seen to be one of the essential parameters that we will need to control
by induction. Note that for the moment, we cannot say whether
the law of the $(\xi_i,\bar Z)$ converges in any sense, as it is not a priori 
clear whether $U_N$ will converge as $N\uparrow \infty$, although 
this would be a natural guess. Note that as far as the computation of the 
marginal probabilities of the Gibbs measures is concerned,
this question is, however, completely irrelevant, in as far as this
term is an even function of the $s_i$. 

\remark It follows from Lemma \ver.3 that
$$
\ln\EQ\exp\left(\sum_{i\in I}\b s_i(\xi_i,\bar Z)\right)=
\frac {\b^2}2|I|\EQ\|\bar Z\|_2^2+O\left(\sfrac 1{\e N}\right)
+R_N
\Eq(G.13bis)
$$
where
$$
\E_{\xi_I}R^2_N\leq \frac CN
\Eq(G.13ter)
$$

\proof The proof of this Lemma relies heavily on the use of the 
Brascamp-Lieb inequalities, Theorem 4.1, 
which are applicable due to our assumptions and Theorem 3.3. 
It was given in [BG1] for $I$ being a single site,
 and we repeat the main steps. 
First note that
$$
\eqalign{
\E_{\xi_I}\EQ e^{\sum_{i\in I}\b s_i (\xi_i,\bar Z)}
&\leq \EQ e^{\frac {\b^2}2\sum_{i\in I} s_i^2  \|\bar Z\|_2^2}\cr
\E_{\xi_I}\EQ e^{\sum_{i\in I}\b s_i (\xi_i,\bar Z)}
&\geq 
\EQ e^{\frac {\b^2}2\sum_{i\in I} s_i^2  \|\bar Z\|_2^2-
\frac {\b^4}4\sum_{i\in I} s_i^4  \|\bar Z\|_4^4}
}
\Eq(P.1)
$$
Note first that if the smallest eigenvalue of  $\nabla^2\Phi\geq \e$, then the 
Brascamp-Lieb inequalities Theorem 4.1 yield
$$
\EQ \|\bar Z\|_2^2\leq \frac M{\e N} +O(e^{-\rho^2N/K})
\Eq(P.2)
$$
and by iterated application
$$
\EQ \|\bar Z\|_4^4\leq 4\frac M{\e^2 N^2} +O(e^{-\rho^2N/K})
\Eq(P.3)
$$
In the bounds \eqv(P.1)  we  now use Corollary 4.2 with $f$ given by 
$\b^2 |I|/2 \|\bar Z\|_2^2$, respectively
by $\b^2 |I|/2 \|\bar Z\|_2^2-\b^4|I|/4 \|\bar Z\|_4^4$ to first move the 
expectation into the exponent, and then \eqv(P.2) and \eqv(P.3) (applied to 
the slightly modified measures $\E_{\Phi_N-tf/N}$, which still retain the 
same convexity properties) to the terms in the exponent. This gives \eqv(G.11).

By very similar computations one shows first that 
$$
\E\left(\EQ e^{\sum_{i\in I}\b s_i (\xi_i,\bar Z)}-
\E_{\xi_I}\EQ e^{\sum_{i\in I}\b s_i (\xi_i,\bar Z)}\right)\leq \frac CN
\Eq(P.4)
$$
Moreover, using again Corollary 4.2, one obtains that
(on the subspace $\bar \O$ where convexity holds)
$$
e^{-\b^2 |I|/2 \frac \a \e }\leq 
\EQ e^{\sum_{i\in I}\b s_i(\xi_i,\bar Z)}
e^{-\b^2 |I|/2 \frac \a \e }  
\Eq(P.5)
$$
These bounds, together with the obvious Lipshitz continuity of the logarithm 
away from zero yield \eqv(G.12). \endproof
 
\remark The above proof follows ideas of the proof of Lemma 4.1 on [T1].
The main difference is that the systematic use of the Brascamp-Lieb 
inequalities that allows us to avoid the appearance of uncontrolled
error terms.    

We now turn to the mean values of the random variables $(\xi_i,\EQ Z)$. 
These are obviously random variables with mean value zero and 
variance $\|\EQ Z\|_2$. Moreover, the variables 
$(\xi_i,\EQ Z)$ and $(\xi_j,\EQ Z)$ are uncorrelated for $i\neq j$. 
Now $\EQ Z$ has one macroscopic component, 
namely the first one, while all others are expected to be small.
 It is thus natural to
expect that these variables will actually converge to a sum of a Bernoulli 
variable $\xi_i^1\EQ Z_1$ plus  
independent gaussians with variance $T_N\equiv\sum_{\mu=2}^M [\EQ Z_\mu]^2$, 
but it is far from trivial to prove this.
It requires in particular at least to show that $T_N$ converges.

We will first prove the following proposition:

\proposition{\ver.4} {\it In addition to our general 
assumption, assume that $\liminf_{N\uparrow\infty}
N^{1/4} T_N=+\infty$, a.s.. 
For $i\in I$, set 
$X_i(N)\equiv \frac 1{\sqrt {T_N}}\sum_{\mu=2}\xi_i^\mu\EQ Z_\mu$. 
Then this family converges to a family of i.i.d. standard normal 
random variables. }                                                        

\remark The assumption on the divergence of $N^{1/4}T_N$ is harmless. 
We will see later that it is certainly
 verified provided $\liminf_{N\uparrow\infty}
N^{1/8} \E T_N=+\infty$.
Recall that our final goal is to approximate (in law) 
 $\sum_{\mu=2}^{M} \xi_i^\mu\EQ Z_\mu$ by    
$\sqrt {T_N} g_i$, where $g_i$ is gaussian. So if 
$T_N\leq N^{-1/4}$, then $\sum_{\mu=2}^{M} \xi_i^\mu\EQ Z_\mu$ is close 
to zero (in law)  
anyway, as is  $\sqrt {T_N} g_i$, and no harm is done if we exchange
the two. We will see that this situation only arises in fact if
$M/N$ tends to zero rapidly, in which  case  all this machinery is not needed.

\proof To prove such a result requires essentially to show that $\EQ Z_\mu$
for all $\mu\geq 2$ tend to zero as $N\uparrow\infty$. 
We note first that by symmetry, for all $\mu\geq 2$,
$\E\EQ Z_\mu=\E\EQ Z_2$. On the other hand,
$$
\sum_{\mu=2}^M [\E\EQ Z_\mu]^2\leq \E \sum_{\mu=2}^M [\EQ Z_\mu]^2 \leq \rho^2
\Eq(G.13)
$$
so that $|\E\EQ Z_\mu|\leq \rho M^{-1/2}$. 

To derive from this a probabilistic bound on $\EQ Z_\mu$ itself we
will use concentration of measure estimates. To do so we need the
following lemma:

\lemma {\ver.5} {\it Assume that $f(x)$ is a  random 
function defined on some open neighborhood $U\subset \R$. 
Assume that $f$ verifies for all $x\in U$
that for all $0\leq r\leq 1$, 
$$
\P\left[|f(x)-\E f(x)|>r\right]\leq c \exp\left(-\frac {Nr^2}c\right)
\Eq(L.69)
$$
and that, at least with probability $1-p$,
 $|f'(x)|\leq C$, $|f''(x)|\leq C<\infty$ both hold uniformly in $U$. 
Then, for any $0<\zeta\leq 1/2$, and for any $0<\d<N^{\zeta/2}$, 
$$
\P\left[|f'(x)-\E f'(x)|> \d N^{-\zeta/2}\right]
\leq \frac {32 C^2}{\d^2}N^\zeta\exp\left(- \frac {\d^4N^{1-2\zeta}}{256c}\right)+p
\Eq(L.69bis)
$$
}

\proof{ Let us assume that $|U|\leq 1$. We may  first assume 
that the boundedness conditions for the derivatives of $f$ 
hold uniformly; by standard arguments one 
shows that if they only hold with probability $1-p$, the effect is nothing more
than the final summand $p$ in \eqv(L.69bis).
The first step in the proof consists in showing that \eqv(L.69)
together with the boundedness of the derivative of $f$ 
implies that $f(x)-\E f(x)$ is uniformly small.
To see this introduce a grid of spacing $\e$, i.e. let
 $U_\e =U\cap \e\Z$. Clearly
$$
\eqalign{
&\P\left[\sup_{x\in U}|f(x)-\E f(x)|>r\right]\cr
&\leq 
\P\Biggl[\sup_{x\in U_\e}|f(x)-\E f(x)|\cr
&\quad\quad+\sup_{x,y: |x-y|\leq \e}
|f(x)-f(y)| +|\E f (x)-\E f(y)|>r\Biggr] \cr
&\leq \P\left[\sup_{x\in U_\e}|f(x)-\E f(x)|>r- 2C\e\right]
\cr
&\leq \e^{-1}  \P\left[|f(x)-\E f(x)|>r- 2C\e\right]
}
\Eq(L.71)
$$
If we choose $\e=\frac r{4C}$, this yields 
$$
\P\left[\sup_{x\in U}|f(x)-\E f(x)|>r\right]\leq 
\frac {4C}r \exp\left(-\frac {N r^2}{4 c}\right)
\Eq(L.72)
$$
Next we show that {\it if } $\sup_{x\in U}|f(x)-g(x)|\leq r$ for two functions 
$f$, $g$ with bounded second derivative, then
$$
|f'(x)-g'(x)|\leq \sqrt {8Cr}
\Eq(L.73)
$$
For notice that
$$
\left|\frac 1\e [f(x+\e)-f(x)]- f'(x)\right|\leq \frac \e 2 
\sup_{x\leq y\leq x+\e} f''(y) \leq C\frac \e 2
\Eq(L.74)
$$
so that
$$
\eqalign{
|f'(x)-g'(x)|&\leq \frac 1\e|f(x+\e) -g(x+\e)-f(x)+g(x)|+ C \e
\cr 
&\leq \frac {2r}\e +C\e
}
\Eq(L.75)
$$
Choosing the optimal $\e=\sqrt{2r/C}$ gives \eqv(L.73). It suffices to combine
\eqv(L.73) with \eqv(L.72) to get
$$
\P\left[|f'(x)-\E f'(x)|>\sqrt {8 rC}\right]\leq \frac {4C}r 
\exp\left(-\frac {N r^2}{4 c}\right)
\Eq(L.76)
$$
Setting $r=\frac {\d^2}{ C N^{\zeta}}$, we arrive at \eqv(L.69bis). \endproof}

We will now use Lemma \ver.5 to control $\EQ Z_\mu$. We define
$$
f(x)=\frac 1{\b N} \ln \int_{B_\rho^{(1,1)}} d^Mz e^{\b N x z_\mu}
e^{-\b N\Phi_{\b,N,M}(z)}
\Eq(L.77)
$$
and denote by $\EQX$ the corresponding modified expectation.
As has by now been shown many times [T1,BG1], $f(x)$ verifies 
\eqv(L.69). Moreover,
$f'(x)= \EQX Z_\mu$ and
$$
f''(x)=\b N \EQX\left(Z_\mu- \EQX Z_\mu\right)^2
\Eq(L.78)
$$
Of course the addition of the linear term to $\Phi$ does not change its 
second derivative, so that we can apply the Brascamp-Lieb inequalities
also to the measure $\EQX$. This shows that 
$$
 \EQX\left(Z_\mu- \EQX Z_\mu\right)^2\leq \frac 1{\e N\b}
\Eq(L.79)
$$ 
which means that $f(x)$ has a second derivative bounded by $c=\frac 1\e$. 


This gives the 

\corollary {\ver.6} {\it There are 
finite positive constants $c, C$ such that, for 
any $0<\zeta\leq \frac 12$, for any $\mu$, 
$$
\P\left[|\EQ Z_\mu-\E\EQ Z_\mu|\geq  N^{-\zeta/2}\right]\leq
  { C}N^\zeta\exp\left(- \frac {N^{1-2\zeta}}{c}\right)
\Eq(L.80)
$$
}

We are now ready to conclude the proof of our proposition. We may choose 
e.g. $\zeta=1/4$ and denote by $\O_N$ the subset of $\O$ where, for all
$\mu$, 
$|\EQ Z_\mu-\E\EQ Z_\mu|\leq  N^{-1/8}$.
Then $\P[\O_N^c]\leq O\left(e^{-N^{1/2}}\right)$.

We will prove the proposition by showing convergence of the 
characteristic function to that of product standard normal distributions, i.e.
we show that for any $t\in \R^I$, $\E\prod_{j\in I} e^{it_jX_j(N)}$ 
converges to $\prod_{j\in I}e^{-\frac 12 t_j^2}$. We have
$$
\eqalign{
\E \prod_{j\in I}e^{it_jX_j(N)}&=
\E_{\xi_{I^c}}\left[\1_{\O_N} \E_{\xi_I} e^{i\sum_{j\in I}t_jX_j(N)} 
+\1_{\O_N^c} \E_{\xi_I}
  e^{i\sum_{j\in I}t_jX_j(N)}\right]\cr 
&=\E_{\xi_{I^c}}\left[\1_{\O_N} \prod_{\mu\geq 2}\prod_{j\in I}
\cos\left(\frac {t_j}{\sqrt {T_N}}
\EQ Z_\mu\right)\right]+O\left(e^{-N^{1/2}}\right)
}
\Eq(L.85)
$$
Thus the second term tends to zero rapidly and can  be forgotten. On the 
other hand, 
on $\O_N$,  
$$
\sum_{\mu=2}^M (\EQ Z_\mu)^4\leq  N^{-1/4}\sum_{\mu=2}^M (\EQ Z_\mu)^2
\leq  N^{-1/4} T_N
\Eq(L.86)
$$ 
Moreover, for any finite $t_j$, for $N$ large enough,
$\left| \frac {t_j}{\sqrt {T_N}}
\EQ Z_\mu\right|\leq 1$. 
Thus, using that   $|\ln \cos x -x^2/2|\leq c x^4$ for $|x|\leq 1$,
and that $$
\eqalign{
&\E_{\xi_{I^c}}\1_{\O_N} \E_\eta e^{i\sum_{j\in I}t_jX_j(N)}\cr
&\leq e^{-\sum_{j\in I}t_j^2/2} 
\sup_{\O_N}\left[\prod_{j\in I}\exp\left(c 
\frac{t_j^4  N^{-1/4}}{T_N}\right) \right] \P_\xi(\O_N)
}
\Eq(L.87)
$$
Clearly, the right hand side 
converges to $e^{-\sum_{j\in I}t_j^2/2}$, provided only that
$N^{1/4}T_N\uparrow \infty$. Since this was assumed, the Proposition is proven.
\endproof

We now control the convergence of our Laplace transform except for the 
three parameters $\mm(N)\equiv \EQ Z_1$, $T_N\equiv \sum_{\mu=2}^M
\left[\EQ Z_\mu\right]^2$ and $U_N\equiv \EQ \|\bar Z\|_2^2$. What we 
have to show is that these quantities converge almost surely and that the 
limits satisfy the equations of the replica symmetric solution of 
Amit, Gutfreund and Sompolinsky  [AGS].

While the issue of convergence is crucial, the technical intricacies of 
its proof are largely disconnected to the question of the convergence of
the Gibbs measures. We will  therefore assume for the moment that these
quantities do converge to some limits and draw the conclusions for the 
Gibbs measures from the results of this section under this assumption (which 
will later be proven to hold). 

Indeed, collecting from Lemma \ver.3 (see the remark
following that lemma) and Proposition \ver.4, we can write
$$
\eqalign{
\mu^{(1,1)}_{\L,\b,\r}[\o]&\left(\{\s_I=s_I\}\right)
=\frac {e^{\b'_N\sum_{i\in I}s_i\left[\mm(N)\xi_i^1+X_i(N)\sqrt {T_N}\right]
+R_N(s_I)}}
{2^I\E_{\s_I}e^{\b'_N\sum_{i\in I}\s_i\left[\mm(N)\xi_i^1+X_i(N)\sqrt {T_N}\right]
+R_N(\s_I)}}
}
\Eq(G.14)
$$
where
$$
\eqalign{
&\b'_N\rightarrow \b\cr
&R_N(s_I)\rightarrow 0\quad \text{in Probability}\cr
&X_i(N)\rightarrow g_i \quad\text {in law}\cr
&T_N\rightarrow \a\rr \quad\text{a.s.}\cr
&\mm(N)\rightarrow \mm \quad\text{a.s.}\cr
}
$$
for some numbers $ \rr,\mm$ and there $\{g_i\}_{i\in \N}$ is a family of 
i.i.d. standard gaussian random variables. 

Putting this together we get that

\proposition{\ver.7} {\it In addition to our general assumptions, 
assume that $T_N\rightarrow \a\rr$, a.s.
and $\mm(N)\rightarrow \mm$, a.s. Then, for any finite $I\subset \N$
$$
\mu^{(1,1)}_{\L,\b,\r}\left(\{\s_I=s_I\}\right)\rightarrow
\prod_{i\in I}\frac {e^{\b s_i
\left[\mm\bar\xi_i^1+g_i\sqrt {\a\rr}\right]}}
{2\cosh\left(\b \s_i\left[\mm\bar\xi_i^1
+g_i\sqrt {\a\rr}\right]\right)}
\Eq(G.15)
$$
where the convergence holds in law with respect to the measure $\P$,
and $\{g_i\}_{\in \in \N}$ is a family of i.i.d. standard normal
random variables and $\{\bar\xi_i^1\}_{i\in \N}$ are independent 
Bernoulli random variables, independent of the $g_i$ and having the 
same distribution as the  variables $\xi_i^1$.
}

To arrive at the convergence in law of the random Gibbs measures,
it is enough to show that \eqv(G.15) holds jointly for any finite
family of cylinder sets, $\{\s_i=s_i,\forall_{i\in I_k}\}, I_k\subset
\N$,
$k=1,\dots,\ell$ (C.f. [Ka], Theorem 4.2).   
But this is easily seen  to hold from the same arguments. Therefore,
denoting by 
$\mu^{(1,1)}_{\infty,\b}$ the random measure
$$
\mu^{(1,1)}_{\infty,\b}[\o](\s)\equiv \prod_{i\in \N}
\frac{e^{\b \s_i[m_1\xi_i^1[\o]+\sqrt{\a\rr}g_i[\o]]}}
{2\cosh\left(\b[m_1\xi_i^1[\o]+\sqrt{\a\rr}g_i[\o]]\right)}
\Eq(G.15bis)
$$
we have

\theo{\ver.8}{\it Under the assumptions of Proposition \ver.7, and
with the same notation,
$$
\mu_{\L,\b,\rho}^{(1,1)}\rightarrow \mu_{\infty,\b}^{(1,1)},
\text{in law, as $\L\uparrow\infty$}, 
\Eq(G.15ter)
$$
}

This result can easily be extended to  the language of metastates.
The following Theorem gives an explicit representation of the 
Aizenman-Wehr metastate in our situation: 

\theo {\ver.9}{ \it Let $\k_\b(\cdot)[\o]$ denote the Aizenman-Wehr
metastate. Under the hypothesis of Proposition \ver.7, for almost all 
$\o$,
for any continuous function $F:\R^k\rightarrow\R$, and cylinder
functions 
$f_i$ on $\{-1,1\}^{I_i}$, $i=1,\dots,k$, one has
$$
\eqalign{
 &\int_{\MM_1(\SS_\infty)}\k_\b(d\mu)[\o] F\left(
\mu(f_1),\dots,\mu(f_k)\right) \cr
&=\int \prod_{i\in I}d\NN(g_i)F\Biggl(\E_{s_{I_1}}f_i(s_{I_1})
\prod_{i\in I_1}\frac {e^{\b \left[ \sqrt {\a\rr}g_i+\mm \xi_i^1[\o]\right]
}}{2\cosh\left(\sqrt {\a\rr}g_i+\mm \xi_i^1[\o]\right)},\dots\cr
&\quad\quad\dots,
\E_{s_{I_k}}f_k(s_{I_k})
\prod_{i\in I_k}\frac {e^{\b \left[ \sqrt {\a\rr}g_i+\mm \xi_i^1[\o]\right]
}}{2\cosh\left(\sqrt {\a\rr}g_i+\mm \xi_i^1[\o]\right)}\Biggr)
}
\Eq(G.16)
$$
where $\NN$ denotes the standard normal distribution.}

\remark Modulo the convergence assumptions, that will be shown to hold in 
the next section, Theorem \ver.9 is the precise statement of Theorem 1.1.
Note that the only difference from Theorem \ver.8 is that the
variables
$\xi_i^1$ that appear here on the right hand side are now the same
as those on the left hand side.

\proof This theorem is proven just as Theorem
\ver.8, except that the ``almost sure version'' of the central limit
theorem, Proposition \ver.4,
which in turn is proven just as Lemma 2.1, is used. The details are
left to the reader.\endproof

\remark Our conditions on the parameters $\a$ and $\b$ place us in the
regime where, according to [AGS] the ``replica symmetry'' is expected to hold.
This is in nice agreement with the remark in [NS4] where 
replica symmetry is linked to the fact that the metastate is 
concentrated on product measures.

\remark
One would be tempted to exploit also the other notions of ``metastate'' 
explained in Section 2. We see that the key to these constructions would be 
an invariance principle associated to the central limit theorem given in 
Proposition \ver.4. However, there are a number of difficulties 
that so far have prevented us from proving such a result. We would have to 
study the random process
$$
X_i^t(N)\equiv \sum_{\mu=2}^{ M(tN)} \xi_i^\mu \E_{\Phi_{tN}} Z_\mu
\Eq(G.100)
$$
(suitably interpolated for $t$ that are not integer multiples of $1/N)$.
If this process was to converge to Brownian motion, 
its increments should converge to independent Gaussians with suitable variance.
But
$$
\eqalign{
X_i^t(N)-X_i^s(N) &= \sum_{\mu=M(sN)}^{ M(tN)} \xi_i^\mu \E_{\Phi_{tN}} Z_\mu
\cr
&+\sum_{\mu=2}^{ M(sN)} \xi_i^\mu 
\left(\E_{\Phi_{tN}} Z_\mu-\E_{\Phi_{sN}} Z_\mu\right)
}
\Eq(G.101)
$$
The first term on the right indeed has the desired properties, 
as is not too hard to check, but the second term is hard to control.

To get some idea of the nature of this process, we recall from 
[BG1,BG2] that $\EQ Z$ is 
approximately given by $c(\b)\frac 1N\sum_{j\in \L\ba I} \xi_j$ 
(in the sense that the 
$\ell_2$ distance between the two vectors is of order $\sqrt\a$ at most).
Let us for simplicity consider only the case $I=\{0\}$. 
If we replace  $\EQ Z$ by this approximation, we are led to study the 
process
$$
Y^t(N)\equiv\frac 1t \sum_{\mu=2}^{\a tN} \xi_0^\mu 
\frac 1N\sum_{i=1}^{tN} \xi_i^\mu
\Eq(G.102)
$$
for $tN, \a t N$ integer and linearly interpolated otherwise.

\proposition {\ver.10} {\it The sequence of processes $Y^t(N)$ defined by 
\eqv(G.102) converges weakly to the gaussian process $ t^{-1}B_{\a t^2}$, 
where $B_s$ is a standard Brownian motion.
} 

\proof 
Notice that  $\xi_0^\mu\xi_i^\mu$ has the same 
distribution as $\xi_i^\mu$, and therefore $Y^t(N)$ has the same distribution
as
$$
\wt Y^t(N)\equiv \frac 1{tN}\sum_{\mu=2}^{\a tN}  \sum_{i=1}^{tN} \xi_i^\mu
\Eq(G.103)
$$
for which the convergence to $B_{\a t^2}$ follows immediately from 
Donsker's theorem.
 \endproof

 
At present we do not see how to extend this result to the real process
of interest, but at least we can expect that some process of this 
type will emerge.

As a final remark we investigate what would happen if we adopted the 
``standard'' notion of limiting Gibbs measures as weak limit points 
along possibly random subsequences. The answer is the following 


\proposition {\ver.10} {\it Under the assumptions of Proposition \ver.7, 
for any finite $I\subset\N$, 
for any $x\in \R^I$, for $\P$-almost all $\o$, there exist
sequences $N_k[\o]$ tending to infinity such that for any $s_I\in \{-1,1\}^I$
$$
\eqalign{
\lim_{k\uparrow\infty}&\mu^{(1,1)}_{N_k,\b}[\o](\{\s_I=s_I\})
\cr
&=\prod_{i\in I}\frac {e^{\b s_i [\mm \xi_i^1[\o]+\sqrt{\a \rr}x_i]}}
{2\cosh(\b [\mm \xi_i^1[\o]+\sqrt{\a \rr}x_i])}
}
\Eq(G.111)
$$
}

\proof To simplify the notation we will write the proof only for the 
case $i=\{0\}$. The general case differs only in notation. 
It is clear that we must show that for almost all $\o$ there exist 
subsequences $N_k[\o]$ such that $X_0(N_k)[\o]$ converges to $x$, 
for any chosen 
value $x$.  Since by assumption $T_N$ converges almost surely to $\a\rr$, it 
is actually enough to show that the variables  
$Y_k\equiv \sqrt {T_{N_k}}X_0(N_k)$ converge to $x$. 
But this follows from the following lemma:

\lemma{\ver.11} {\it Define $Y_k\equiv \sqrt {T_{N_k}}X_0(N_k)$. 
For any $ x\in \R^I$ and any $\e>0$,
$$
\P\left[ Y_k \in (x_0-\e,x_0+\e) \,\,\hbox{i.o.}\,\right]=1
\Eq(G.112)
$$
}

\proof Let us denote by $\FF_\xi$ the sigma algebra generated by the random 
variables $\xi_i^\mu, \mu\in \N, i\geq 1$. Note that
$$
\P\left[ X_k \in (x_0-\e,x_0+\e) \,\,\hbox{i.o.}\,\right]=\E\left(
\P\left[ X_k \in (x_0-\e,x_0+\e) \,\,\hbox{i.o.}\,\mid\FF_\xi
\right]\right)
\Eq(G.113) 
$$
so that it is enough to prove that for almost all
$\o$,
$\P\left[ X_k \in (x_0-\e,x_0+\e) \,\,\hbox{i.o.}\,\mid\FF_\xi
\right]=1$. 

Let us define the random variables 
$$
\wt Y_k\equiv \sum_{\mu=M( N_{k-1})+1}^{M(N_k)}
\xi_0^\mu \E_{\Phi_{N_k}}Z_\mu
\Eq(G.114)
$$
Note first that
$$
\E\left(Y_k-\wt Y_k\right)^2=\E\sum_{\mu=2}^{M(N_{k-1})}
\left(\E_{\Phi_{N_k}}Z_\mu\right)^2
\leq M(N_{k-1})\E\left(\E_{\Phi_{N_k}}Z_2\right)^2\leq 
\rho^2\frac {N_{k-1}}{N_k}
\Eq(G.115)
$$ 
Thus, if $N_k$ is chosen such that 
$\sum_{k=1}^\infty
\frac {N_{k-1}}{N_k}<\infty$, by the first Borel-Cantelli lemma,
$$
\lim_{k\uparrow\infty} (Y_k-\wt Y_k) =0\text {a.s.}
\Eq(G.116)
$$
On the other hand, the random variables $\wt Y_k$ are conditionally 
independent,
given $\FF_\xi$. Therefore, by the second Borel-Cantelli lemma
$$
\P\left[ X_k \in (x_0-\e,x_0+\e) \,\,\hbox{i.o.}\,\mid\FF_\xi
\right]=1
\Eq(G.117)
$$
if 
$$
\sum_{k=1}^\infty \P\left[ X_k \in (x_0-\e,x_0+\e)\,\mid\FF_\xi
\right]=\infty
\Eq(G.118)
$$
But for almost all $\o$, $\wt Y_k$ conditioned on $\FF_\xi$ 
converges to a gaussian of variance $\a\rr$ (the proof is identical to that 
of Proposition \ver.3), so that for almost all $\o$, as $k\uparrow\infty$
$$
 \P\left[ X_k \in (x_0-\e,x_0+\e)\,\mid\FF_\xi
\right]\rightarrow \frac 1{\sqrt {2\pi\a\rr}}
\int_{x-\e}^{x+\e}d y e^{-\frac{y^2}{2\a\rr}}>0
\Eq(G.119)
$$
which implies \eqv(G.118) and hence \eqv(G.117). Putting this together with
\eqv(G.116) concludes the proof of the lemma, and of the proposition. \endproof

Some remarks concerning the implications of this proposition are in place. 
First, it shows that if the standard definition of limiting Gibbs measures
as weak limit points is adapted, then we have discovered that in the 
Hopfield model all product measures on $\{-1,1\}^\N$ are extremal 
Gibbs states. Such a statement contains some information, but it is clearly 
not useful as information on the approximate nature of a finite volume 
state. This confirms our discussion in Section 2 on the necessity to 
use a metastate formalism.

Second, one may ask whether conditioning or the application 
of external fields of vanishing strength as discussed in Section 2 can 
improve the convergence behaviour of our measures. The answer appears
 obviously to be no. Contrary to a situation where a symmetry is present 
whose breaking biases the system to choose one of the possible states, the 
application of an arbitrarily weak field cannot alter anything. 

Third, we note that the 
total set of limiting Gibbs measures does not depend on the 
conditioning on the ball $B_\rho^{(1,1)}$, while the metastate obtained does 
depend on it. Thus the conditioning allows us to construct two metastates 
corresponding to each of the stored patterns. These metastates are in a 
sense extremal, since they are concentrated on the set of extremal 
(i.e. product) measures of our system. Without conditioning one can 
construct other metastates (which however we cannot control explicitly in our 
situation).

\newpage

\def\EQ{\E_{\Phi_{N}}}
\def\EQQ{\E_{\Phi_{N+1}}
\def\EQX{\E_{\Phi_N,x}}}
\chap{6. Induction and the replica symmetric solution}6

We now conclude our analysis by showing that the quantities 
$U_N\equiv\EQ \|\bar Z\|_2^2$, $\mm(N)\equiv\EQ Z_1$ 
and $T_N\equiv\sum_{\mu=2}^M [\EQ Z_\mu]^2$ actually do converge  
almost surely under 
our general assumptions. The proof consist of two steps: First we show that
these quantities are self-averaging and then the convergence of their 
mean values is proven by induction. We will assume throughout this section 
that the parameters $\a$ and $\b$ are such that local convexity holds. 
We stress that this section is entirely based on ideas of Talagrand [T1]
and Pastur, Shcherbina and Tirozzi [PST] and is mainly added for the 
convenience of the reader.                             

Thus our first result will be:
 
\proposition {\ver.1} {\it Let $A_N$ denote any of the three 
quantities $U_N$, $\mm(N)$ or $T_N$. Then
  there are 
finite positive constants $c, C$ such that, for 
any $0<\zeta\leq \frac 12$,  
$$
\P\left[|A_N-\E A_N|\geq  N^{-\zeta/2}\right]\leq
  { C}N^\zeta\exp\left(- \frac {N^{1-2\zeta}}{c}\right) 
\Eq(G.20)
$$
}

\proof The proofs of these three statements are all very similar 
to that of Corollary 5.6. Indeed, for $\mm(N)$, \eqv(G.20)
is a special case of that corollary. In the two other cases, we just need to 
define the  appropriate analogues of the `generating function' $f$ from
\eqv(L.77). They are
$$
g(x)\equiv \frac 1{\b N}\ln 
 \EQ\EQ'     e^{\b N x (\bar Z, \bar Z')}
\Eq(L.82)
$$
in the case of $T_N$ and
$$
\tilde g(x)\equiv \frac 1{\b N}\ln 
 \EQ\EQ'     e^{\b N x\|\bar Z\|_2^2}
\Eq(L.83)
$$
The proof then proceeds as in that of Corollary \ver.6. We refrain from 
giving the details. \endproof

We now turn to the induction part of the proof and derive a recursion 
relation for the three quantities above. In the sequel it will be 
convenient to introduce a site $0$ that will replace the set $I$ and 
to  set   $\xi_0=\eta$.  
Let us define
$$
u_N(\t)\equiv \ln \EQ e^{\b\t(\eta,Z)} 
\Eq(I.1)
$$
We also set 
$
v_N(\t)\equiv \t\b(\eta,\EQ Z)     
$ and $w_N(\t)\equiv u_N(\t)-v_N(\t)$. 
In the sequel we will need the following auxiliary result

\lemma {\ver.2} {\it 
Under our general assumptions
\item{(i)}
$\frac 1{\b\sqrt {T_N}}\frac d{d\t} v_N(\t)$ converges weakly to 
a standard gaussian random variable. 
\item{(ii)} 
$\left|\frac d{d\t} w_N(\t)-\t \b^2\E\EQ\|\bar Z\|_2^2\right|$
converges to zero in probability.
 }

\proof (i) is obvious from Proposition 5.4 and the definition of $v_N(\t)$. 
To prove (ii), note that
$w_N(\t)$ is convex and $\frac {d^2}{d\t^2} w_N(\t)\leq \frac {\b\a}{\e}$. 
Thus, {\it if} $\hbox{var}\left(w_N(\t)\right)\leq \frac C{\sqrt {N}}$, then 
 $\hbox{var}\left(\frac {d}{d\t} w_N(\t)\right)\leq \frac {C'}{ N^{1/4}}$
by  a standard result similar in spirit to Lemma 5.5 (see e.g. 
[T2], Proposition 5.4).  On the other hand, 
$|\E w_N(\t)-\frac {\t^2\b^2}2\E\EQ\|\bar Z\|_2^2|\leq \frac K{\sqrt N}$,
by Lemma 5.3, which, together with the boundedness of the second 
derivative of $w_N(\t)$ implies that 
$|\frac d{d\t} \E w_N(\t)-\t\b^2
\E\EQ\|\bar Z\|_2^2|\downarrow 0$. This 
means that  $\hbox{var}\left(w_N(\t)\right)\leq \frac C{\sqrt {N}}$
implies the lemma. Since we already know from G.11ter) 
that $\E R_N^2\leq \frac KN$, it is
enough to prove $\hbox{var}\left(\EQ\|\bar Z\|_2^2\right)\leq 
\frac C{\sqrt {N}}$. This follows just as the corresponding concentration 
estimate for $U_N$. \endproof

 
We are now ready to start the induction procedure. We will place ourselves on 
a subspace $\wt\O\subset \O$ where for  all but finitely many $N$
$|U_N-\E U_N|\leq N^{-1/4}$, $|T_N-\E T_N|\leq N^{-1/4}$, etc. 
This subspace has  probability one by our estimates. 

Let us note that by (iii) of Proposition 3.1, $\EQ Z_\mu$ and 
$\int d\QQ_{N,\b,\rho}^{(1,1)} (m) m_\mu$ differ only by an exponentially small
term. Thus 
$$
\EQ Z_\mu = \frac 1N \sum_{i=1} \xi_i^\mu 
\int \mu_{N,\b,\rho}^{(1,1)}(d\s) \s_i +O\left(e^{-cM}\right)
\Eq(I.2)
$$
and, by symmetry,
$$
\E\EQQ(Z_\mu)=\E \eta^\mu \int \mu_{N+1,\b,\rho}^{(1,1)}(d\s) \s_0 +
O\left(e^{-cM}\right)
\Eq(I.3)
$$
Using Lemma 5.2 and the definition of $u_N$, this gives
$$
\E\EQQ(Z_\mu)= \E\eta^\mu\frac {e^{u_N(1)}-e^{u_N(-1)}}{e^{u_N(1)}+e^{u_N(-1)}}
+O\left(e^{-cM}\right)
\Eq(I.4)
$$
where to be precise one should note that the left and right hand side 
are computed at temperatures $\b$ and $\b'=\frac N{N}\b$, respectively,
and that the value of $M$ is equal to $M(N+1)$ on both sides;
that is, both sides correspond to slightly different 
values of $\a$ and $\b$, but we will see that this causes no problems.

Using our concentration results and  Lemma 5.3  this gives
$$
\E\EQQ(Z_\mu)=\E\eta^\mu \tanh\left(\b(\eta^1 \E \mm(N)+ \sqrt{\E T_N}X_0(N))
\right) +O(N^{-1/4})
\Eq(I.5)
$$
Using further Proposition 5.4 we get a first recursion for $\mm(N)$:
$$
\mm(N+1)=\int d\NN(g) \tanh\left(\b(\E \mm(N)+ \sqrt{\E T_N}g)
\right) +o(1)
\Eq(I.6)
$$
 
\remark The error term in \eqv(I.6) can be sharpened to $O(N^{-1/4})$ by using 
instead of Lemma 5.3 a trick, attributed to Trotter,
that we learned from Talagrand's paper [T1]
(see the proof of Proposition 6.3 in that paper).  

We need of course a recursion for $T_N$ as well. From here on there is 
no great difference from the procedure in [PST], except that 
the $N$-dependences have to be kept track of carefully. This was outlined in 
[BG4] and we repeat the steps for the convenience of the reader. 
To simplify the notation, we ignore all the $O(N^{-1/4})$
error terms and put them back in the end only. Also, the remarks concerning
$\b$ and $\a$ made above apply throughout.

Note that 
$T_N= \|\EQ Z\|_2^2 -(\EQ Z_1)^2$ and 
$$
\eqalign{ 
\E\|&\EQQ Z\|_2^2
=\sum_{\mu=1}^M\E\left(\frac 1{N+1}\sum_{i=0}^{N}\xi_i^\mu
\mu_{\b,{N+1},M}(\s_i)\right)^2\cr
&=\frac {M} {N+1} \E  \left(\mu^{(1,1)}_{\b,N+1,M}(\s_0)\right)^2\cr
&+\sum_{\mu=1}^M \E \xi_{0}^\mu \mu^{(1,1)}_{\b,N+1,M}(\s_{0})
\left(\frac 1{N+1} \sum_{i=1}^N \xi_i^\mu \mu_{\b,{N+1},M}(\s_i)\right)
}
\Eq(L.912)
$$
Using Lemma 5.2 as in the step leading to \eqv(I.4), we get for the first 
term in \eqv(L.912)
$$
 \E  \left(\mu^{(1,1)}_{\b,N+1,M}(\s_0)\right)^2
= \E \tanh^2\left(\b(\eta_1 \EQ Z_1 + \sqrt{\E T_{N}})\right)\equiv \E Q_{N}
\Eq(L.911)
$$
For the second term, we use the identity from [PST]
$$
\eqalign{
\sum_{\mu=1}^M &\xi_{0}^\mu\left(\frac 1{N}
 \sum_{i=1}^N \xi_i^\mu \mu_{\b,{N+1},M}(\s_i)\right)
= \frac {\sum_{\s_0}\EQ (\xi_0,X) e^{\b\s_0 (\xi_0,X)}}
        {\sum_{\s_0}\EQ e^{\b\s_0 (\xi_0,X)}}\cr
=&\b^{-1}\frac{\sum_{\t=\pm 1} u_N{}'(\t) e^{u_N(\t)}}{\sum_{\t=\pm 1}
e^{u_N(\t)}}
}
\Eq(L.913)
$$ 
Together with Lemma \ver.2 one concludes that in law up to small 
errors
$$
\eqalign{
&\sum_{\mu=1}^M \xi_{0}^\mu\left(\frac 1{N+1}
 \sum_{i=1}^N \xi_i^\mu \mu_{\b,{N+1},M}(\s_i)\right)
=\xi_{0}^1\EQ Z_1 +\sqrt{\E T_N} X_N \cr
&\quad+\b \EQ\|\bar Z\|_2^2
\tanh\b\left(\xi_{0}^1\EQ Z_1 +\sqrt{\E T_N} X_N\right)
}
\Eq(L.914)
$$
and so
$$ 
\eqalign{  
\E\|\EQQ Z\|_2^2
 &=\a \E Q_{N} +\E \Biggl[ \tanh\b\left(\xi_{0}^1\EQ Z_1 +
\sqrt{\E T_N} X_N\right) \cr
&\times\left[\xi_{0}^1\EQ Z_1 +\sqrt{\E T_N} X_N\right]\Biggr]\cr
&+\b\E \EQ\|\bar Z\|_2^2  
\tanh^2\b\left(\xi_{0}^1\EQ Z_1 +
\sqrt{\E T_N} X_N\right)
}
\Eq(L.915)
$$
Using the self-averaging properties of $\EQ\|\bar Z\|_2^2$,
the last term is of course essentially equal to 
$$
\b\E \EQ\|\bar Z\|_2^2 \E Q_{N}
\Eq(L.916)
$$
The appearance of $ \EQ\|\bar Z\|_2^2$ is disturbing, as it
introduces a new quantity into the system. Fortunately, it is the last one.
The point is that proceeding as above, we can show that
$$
\eqalign{
\E \EQQ\|Z\|_2^2=&\a+\E \Biggl[ \tanh\b\left(\xi_{N+1}^1
\EQ Z_1 +
\sqrt{\E T_N} X_N\right) \cr
&\times\left[\xi_{0}^1\EQ Z_1 +\sqrt{\E T_N} X_N\right]\Biggr]
+\b\E \EQ\|\bar Z\|_2^2 \E Q_{N}
}
\Eq(L.917)
$$
so that setting $U_{N}\equiv \EQ\|\bar Z\|_2^2$, we get, subtracting 
\eqv(L.915) from \eqv(L.917), the simple 
recursion
$$
\E U_{N+1} =\a(1-\E Q_N)+\b(1-\E Q_N)\E U_N
\Eq(L.918)
$$
From this we get (since all quantities considered are self-averaging, we drop
the $\E $ to simplify the notation), setting  $\mm(N)\equiv \EQ Z_1$,
$$
\eqalign{
 T_{N+1} &=- ( \mm(N+1))^2 +\a Q_N  +\b U_N Q_N
\cr
&+\int d\NN(g)[\mm(N) +\sqrt {T_N}g]
\tanh\b(\mm(N)+  \sqrt {T_N}g)
\cr
&= \mm(N+1)(\mm(N)-\mm(N+1)) +\b U_N Q_N + \b T_N (1-Q_N)+\a Q_N
}
\Eq(L.919)
$$
where we used integration by parts.
The complete system of recursion relations can thus be written as
$$
\eqalign{
\mm(N+1)&=\int d\NN(g) \tanh \b\left(\mm(N) +\sqrt{T_N} g\right)+O(N^{-1/4})\cr
 T_{N+1} & = \mm(N-1)(\mm(N)-\mm(N+1)) +\b U_N Q_N + \b T_N (1-Q_N)+
\a Q_N+O(N^{-1/4})\cr
 U_{N+1} &=\a(1- Q_N)+\b(1- Q_N) U_N+O(N^{-1/4})\cr
Q_{N+1} &= \int  d\NN(g) \tanh^2 \b\left(  \mm(N) +\sqrt{ T_N} g\right)
+O(N^{-1/4})\cr
}
\Eq(L.920)
$$
If the solutions to this system of equations converges, than the limits 
$\rr=\lim_{N\uparrow \infty} T_N/\a$, $\qq=\lim_{N\uparrow \infty} Q_N$ and
$\mm=\lim_{N\uparrow \infty}\mm(N)$ ($u\equiv\lim_{N\uparrow \infty} U_N$
can be eliminated) must satisfy the equations
$$
\mm =\int d\NN (g) \tanh(\b( \mm +\sqrt{\a \rr}g))
\Eq(L.RS1)
$$
$$
\qq=\int d\NN (g)  \tanh^2(\b( \mm +\sqrt{\a \rr}g))
\Eq(L.RS2)
$$
$$
\rr=\frac \qq{(1-\b+\b \qq)^2}
\Eq(L.RS3)
$$
which are the equations for the replica symmetric solution of the 
Hopfield model found by Amit et al. [AGS]. 

In principle one might think that to prove convergence it is enough to study 
the stability of the dynamical system above without the error 
terms. However, this is not quite true. Note that the parameters 
$\b$ and $\a$ of the 
quantities on the two sides of the equation differ slightly
(although this is suppressed in the notation). In particular, 
if we iterate too often, $\a$ will tend to zero. 
The way out of this difficulty was proposed by Talagrand [T1]. We will 
briefly explain his idea. In a simplified notation, we are in the following 
situation: We have a sequence $X_n(p)$ of functions depending 
on a parameter $p$. There is an explicit sequence $p_n$, satisfying
$|p_{n+1}-p_n|\leq c/n$ and a functions $F_p$ such that
$$
X_{n+1}(p_{n+1}) =F_{p_n}(X_n(p_n)) +O(n^{-1/4})
\Eq(I.7)
$$
In this setting, we have the following lemma.

\lemma{\ver.3} {\it Assume that there exist a domain $D$ containing a single 
fixed point $X^*(p)$ of $F_p$. Assume that $F_p(X)$ is Lipshitz continuous
as a function of $X$, Lipshitz continuous  as a function of $p$ 
uniformly for $X\in D$   and that for all $X\in D$,
$F^n_p(X) \rightarrow X^*(p)$. Assume we know that for all $n$ large enough,
$X_n(p)\in D$. 
Then 
$$
\lim_{n\uparrow\infty}X_n(p)=X^*(p)
\Eq(I.8)
$$
} 

\proof Let us choose a integer valued monotone 
increasing function $k(n)$ such that 
$k(n)\uparrow \infty $ as $n$ goes to infinity. Assume e.g. $k(n)\leq \ln n$.
We will show that 
$$
\lim_{n\uparrow\infty}X_{n+k(n)}(p)=X^*(p)
\Eq(I.9)
$$
 To see this, note first that
$|p_{n+k(n)}-p_n|\leq \frac {k(n)}n$. By \eqv(I.7), we have that
using the Lipshitz properties of $F$
$$
X_{n+k(n)}(p) =F^{k(n)}_{p}(X_n(p_n)) +O(n^{-1/4}) 
\Eq(I.10)
$$
where we choose $p_n$ such that $p_{n+k(n)}=p$. 
Now since $X_n(p_n)\in D$, 
$
\left|F^{k(n)}_p(X_n(p_n)-X^*(p)\right|\downarrow 0$
as $n$ and thus $k(n)$ goes to infinity, so that \eqv(I.10) implies 
\eqv(I.9). But \eqv(I.9) for any slowly diverging function
$k(n)$ implies the convergence of $X_n(p)$, as claimed.
\endproof

This lemma can be applied to the recurrence \eqv(L.919). The main point to 
check is whether the corresponding $F_\b$ attracts a domain in which the
parameters $\mm(N), T_N, U_N, Q_N$  are a priori located due tho the 
support properties of the measure $\wt\QQ_{N,\b,\rho}^{(1,1)}$. This stability 
analysis was carried out (for an equivalent system) by Talagrand and answered
 to the affirmative. We do not want to repeat this tedious, but in principle 
elementary computation here. 

We would like to make, however, some remarks. It is clear that if we 
consider conditional measures, then we can always force the 
parameters $\mm(N), R_N, U_N, Q_N$ to be in some domain. Thus, in principle, 
we could first study the fixpoints of \eqv(L.919), determine their
domains of attraction and then define corresponding 
conditional Gibbs measures. However, these measures may then be metastable.
Also, of course, at least in our derivation, do we need to verify the 
local convexity in the corresponding domains since this was used in the 
derivation of the equations \eqv(L.919).

\newpage

\frenchspacing
\chap{References}4
\item{[AGS]} D.J. Amit, H. Gutfreund and H.
Sompolinsky, ``Statistical mechanics of neural networks near saturation'',
Ann. Phys. {\bf 173}, 30-67 (1987).
 \item{[AW]}  M. Aizenman, and J. Wehr, ``Rounding effects on
quenched randomness on first-order phase transitions'',
Commun. Math. Phys. {\bf 130}, 489 (1990).
643-664 (1993).
\item{[BG1]} A. Bovier and V. Gayrard, ``The retrieval phase of the 
Hopfield model, A rigorous analysis of the overlap distribution'', 
Prob. Theor. Rel. Fields {\bf 107}, 61-98 (1997).
\item{[BG2]}  A. Bovier and V. Gayrard, ``The Hopfield model as a 
generalized random mean field model'', in ``Mathematics of spin glasses and 
neural  networks'', A. Bovier and P. Picco, Eds., Progress in Probablity,
Birkh\"auser, Boston, (1997).
\item{[BG3]} A. Bovier and V. Gayrard, ``An almost sure central limit
theorem for the Hopfield model'', to appear in Markov Proc. Rel. Fields (1997).
\item{[BGP1]} A. Bovier, V. Gayrard, and P. Picco, ``Gibbs states
of the Hopfield model in the regime of perfect  memory'',
Prob. Theor. Rel. Fields {\bf 100}, 329-363 (1994).
\item {[BGP2]} A. Bovier, V. Gayrard, and P. Picco,
``Gibbs states of the Hopfield model with extensively many patterns'',
 J. Stat. Phys. {\bf 79}, 395-414 (1995).
\item{[BGK]} A. Bovier, V. Gayrard, and Ch. K\"ulske, in preparation.
\item{[BK]}  A. Bovier and Ch. K\"ulske, A rigorous renormalization group
method for interfaces in random media,
Rev. Math. Phys. {\bf 6}, 413-496 (1994).
\item{[BL]} H.J. Brascamp and E.H. Lieb, ``On extensions of the 
Brunn-Minkowski and P\'ekopa-Leindler theorems, 
including inequalities for log concave functions, and with an application to 
the diffusion equation'', J. Funct. Anal. {\bf 22}, 366-389 (1976).
\item{[H]} B. Helffer, ``Recent results and open problems on Schr\"odinger 
operators, Laplace integrals, and transfer operators in large dimension'' 
(1996). 
\item{[HH]} P. Hall and C.C. Heyde, ``Martingale limit theory
and its applications'', Academic Press, New York (1980).
\item{[Ho]} J.J. Hopfield, ``Neural networks and physical systems
with emergent collective computational abilities'', Proc. Natl.
Acad. Sci. USA {\bf 79}, 2554-2558 (1982).
\item{[HS]} B. Helffer and J. Sj\"ostrand, ``On the correlation for Kac-like
models in the convex case'', J. Stat. Phys. {\bf 74}, 349-409 (1994).
\item{[Ka]} O. Kallenberg, ``Random measures'', 
Academic Press, New York (1983).
\item{[K]} Ch. K\"ulske, ``Metastates in disordered mean field models:
random field and Hopfield models'', to appear in J. Stat. Phys. (1997).
\item{[MPV]} M. M\'ezard, G. Parisi, and M.A. Virasoro, 
``Spin-glass theory
and beyond'', { World Scientific}, Singapore (1988).
\item{[N]} Ch. Newman, ``Topics in disordered systems'', 
     Birkh\"auser, Boston (1997).
\item{[NS1]}  Ch.M. Newman and D.L. Stein, ``Multiple states and the 
thermodynamic limits in short ranged Ising spin glass models'', Phys. Rev. 
{\bf B 72}, 973-982 (1992).
\item {[NS2]} Ch.M. Newman and D.L. Stein, ``Non-mean-field behaviour in
realistic spin glasses'', Phys. Rev. Lett. {\bf 76}, 515-518 (1996).
\item{[NS3]} Ch.M. Newman and D.L. Stein, 
``Spatial inhomogeneity and thermodynamic chaos'', Phys. Rev. Lett. {\bf 76},
4821-4824 (1996).
\item{[NS4]}  Ch.M. Newman and D.L. Stein, 
``Thermodynamic chaos and the structure of short range
spin glasses'', in ``Mathematical aspects of spin glasses and neural 
networks'', A. Bovier and P. Picco (Eds.), Progress in Probability,
Birkh\"auser, Boston (1997).
\item{[NS5]} C.M. Newman and D.L. Stein, ``Ground state structure in a highly 
disordered spin glass model'', J. Stat. Phys. {\bf 82}, 1113-1132 (1996).
 \item{[PST]} L. Pastur, M. Shcherbina, and B. Tirozzi, ``The replica
 symmetric solution without the replica trick for the Hopfield model'',
 J. Stat. Phys. {\bf 74}, 1161-1183 (1994).
\item{[T1]} M. Talagrand, ``Rigorous results for the Hopfield model with many
 patterns'', preprint 1996, to appear in Probab. Theor. Rel. Fields.
\item{[T2]} M. Talagrand, ``The Sherrington-Kirkpatrick model: 
A challenge for 
mathematicians'', preprint 1996, to appear in Prob. Theor. Rel. Fields.

\end